\documentclass[lettersize,journal]{IEEEtran}
\usepackage{amstext}
\usepackage{amssymb}
\usepackage{coseoul}
\usepackage{modular}
\usepackage{babel}
\usepackage{breqn}
\usepackage{bm}
\usepackage[pdfpagemode=UseNone,pdfstartview=FitH,colorlinks=true,linkcolor=blue,citecolor=blue,urlcolor=blue]{hyperref}
\usepackage[all]{hypcap}
\usepackage[T1]{fontenc}
\usepackage[latin9]{inputenc}
\usepackage{amsmath,amsfonts}
\usepackage{algorithmic}
\usepackage{algorithm}
\usepackage{array}
\usepackage[caption=false,font=normalsize,labelfont=sf,textfont=sf]{subfig}
\usepackage{textcomp}
\usepackage{stfloats}
\usepackage{url}
\usepackage{verbatim}
\usepackage{graphicx}
\usepackage{hyperref}

\usepackage{caption}
\usepackage{subcaption}
\usepackage[caption=false]{subfig}
\usepackage{tikz}

\renewcommand{\baselinestretch}{1.5}
\newcommand{\abs}[1]{\left \lvert #1 \right \rvert}

\begin{document}

\title{On the electromagnetic couplings in superconducting qubit circuits}
\author{Ebrahim Forati, \IEEEmembership{Senior member, IEEE},  Brandon W. Langley, Ani Nersisyan, and Reza Molavi \IEEEmembership{member, IEEE}
%\thanks{This manuscript was received in April 2024}
\thanks{Authors are with Google Quantum AI, Goleta, CA 93117 USA (e-mail: forati@google.com). }}

%\markboth{IEEE Transactions on Microwave Theory and Techniques}%

%\IEEEpubid{0000--0000~\copyright~2023 IEEE}
\maketitle

\begin{abstract}
The precise engineering of electromagnetic couplings is paramount for constructing scalable and high-fidelity superconducting quantum processors. While essential for orchestrating qubit operations, these couplings also present significant design challenges, including the mitigation of crosstalk and the management of environmental decoherence. A clear and unified theoretical framework is therefore crucial for the design, simulation, and analysis of these complex quantum circuits. This paper presents a comprehensive theoretical treatment of the fundamental electromagnetic coupling mechanisms in superconducting devices. Starting from first principles, we formulate the equations of motion and derive the input-output relations for canonical systems, including a single resonator coupled to a multi-port microwave network, interacting resonators, and coupled transmission lines. We review rigorous definitions for key parameters such as the energy decay rate ($\kappa$) and the dimensionless coupling coefficient ($\zeta$) and connect these formalisms to practical methods of parameter extraction from electromagnetic simulations. This work provides a rigorous and pedagogical foundation for understanding and modeling linear electromagnetic interactions, serving as a vital resource for the development of advanced superconducting quantum hardware.
\end{abstract}


\section{Introduction}

Superconducting circuits have emerged as a leading platform for building fault-tolerant quantum computers, enabling groundbreaking demonstrations of quantum supremacy \cite{arute2019quantum} and error correction \cite{krinner2022realizing, kjaergaard2020superconducting}. The success of this modality is built upon the ability to precisely control coherent quantum states through meticulously engineered electromagnetic interactions \cite{kjaergaard2020superconducting}. Foundational quantum operations---from single-qubit rotations and multi-qubit entangling gates to high-fidelity projective measurements---are all mediated by carefully designed couplings between circuit elements such as qubits, resonators, and transmission lines. The theoretical framework of circuit quantum electrodynamics (cQED) provides a powerful lens for understanding these interactions, where superconducting circuits behave as artificial atoms coupled to microwave photons in on-chip resonators \cite{blais2021circuit}.

The precise control of these couplings is a central theme in modern quantum hardware development. On one hand, significant innovation in hardware, such as the design of tunable couplers, has enabled precise control over qubit interactions. These couplers can dynamically mediate or nullify coupling, reducing parasitic ZZ crosstalk and mitigating frequency crowding in multi-qubit processors \cite{sung2021realization, mundada2019suppression}. The ability to engineer these strong, controllable interactions is crucial for performing the fast, high-fidelity two-qubit gates that are a critical ingredient for any quantum algorithm \cite{xu2020high}. On the other hand, unwanted couplings to other quantum systems or the electromagnetic environment are a primary source of decoherence and computational errors. Spurious crosstalk between neighboring components remains a fundamental obstacle to scaling, as it can lead to correlated errors that are particularly detrimental to quantum error correction codes \cite{mundada2019suppression, zhao2022high}. Consequently, sophisticated techniques for engineering the electromagnetic environment, such as the use of Purcell filters to protect qubits from radiative decay through their readout resonators, are essential for achieving long coherence times alongside rapid measurements \cite{heinsoo2018rapid}.

While the literature is rich with advanced device designs and experimental breakthroughs, it often presumes a deep, intuitive understanding of the underlying theoretical models of electromagnetic coupling. Foundational concepts like the rotating-wave approximation (RWA) are ubiquitously applied, yet the conditions for their validity and the consequences of their breakdown require careful consideration \cite{sank2024balanced}. For researchers and engineers entering the field, or for those seeking to bridge the gap between abstract models and physical device implementation, a consolidated and pedagogical resource that systematically derives and unifies the various analytical formalisms is invaluable.

Generally, designing superconducting quantum chips requires a set of linear and nonlinear resonators along with a number of input/output transmission lines, regarded as an open quantum mechanical system. The Hamiltonian of such a system can be quantified in several ways, among which the Lumped Oscillator Model (LOM) is attractive. Extracting the precise linear equivalent circuit of the physical design is a crucial step in this approach. This paper aims to fill this need by providing a clear, first-principles-based exposition of the electromagnetic couplings that form the building blocks of modern superconducting quantum circuits. We seek to formalize the connection between different, yet complementary, descriptions of circuit behavior, from lumped-element models to the fields of distributed resonators and the input-output theory of open quantum systems. By establishing a unified mathematical framework, this work serves as a foundational reference for the analysis, design, and simulation of high-performance superconducting quantum devices. The manuscript is organized as follows:

 \begin{enumerate}
     \item We begin by analyzing a single resonator coupled to a transmission line, establishing the fundamental concepts of mode amplitudes and the energy decay rate, $\kappa$.
     \item  This analysis is then generalized to a resonator coupled to a multi-port microwave network, providing a versatile tool for modeling complex circuit environments.
     \item Next, we investigate the time dynamics of two coupled resonators, deriving the unitless coupling coefficient, $\zeta$, and describing the hybridization of modes and the phenomenon of avoided crossing.
     \item Finally, we review the analysis of coupled transmission lines, which are ubiquitous as directional couplers and control lines in readout and qubit architectures.
 \end{enumerate}
   Throughout these sections, our goal is to provide a clear and rigorous derivation of the equations of motion and to connect the theoretical parameters to methods for their practical extraction from numerical solvers and experimental measurements.

\section{A single resonator}

\begin{figure}[t]
\begin{centering}
\includegraphics[width=5cm]{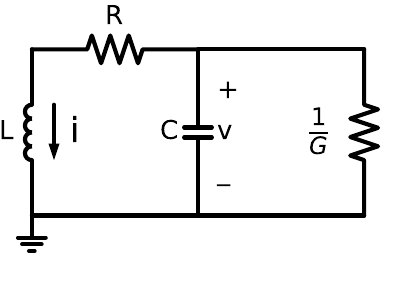}
\caption{An LC resonator.\label{fig:Single-LC-resonator}}
\end{centering}
\end{figure}

Consider a lossless LC resonator with current and voltage definitions shown in Fig. \ref{fig:Single-LC-resonator}, ignoring R and G for the initial discussion. 
Using Kirchhoff's laws, and defining $\omega_0=1/\sqrt{\mathrm{LC}}$, 
\begin{equation}
  \frac{d^2v}{dt^2}+\omega_{0}^{2}v=0,\label{eq:3}
\end{equation}
or
\begin{equation}
  \frac{d^2i}{dt^2}+\omega_{0}^{2}i=0,\label{eq:4}
\end{equation}
where $v$ and $i$ are real numbers with different initial conditions. 
Alternatively, we may define \cite{haus1984waves}
\begin{align}
    a_{\pm} &= \frac{1}{\sqrt{2 \omega_0  Z }}\left(v\pm j Z i\right),  \label{eq:12}
%   a_{+} &= \left(\frac{C}{4L}\right)^{1/4}\left(v+j\sqrt{\frac{L}{C}}i\right),  \label{eq:12} \\
%   a_{-} &= \left(\frac{C}{4L}\right)^{1/4}\left(v-j\sqrt{\frac{L}{C}}i\right), \label{eq:13}
% \end{align}
\end{align}
where $Z=\sqrt{\mathrm{L/C}}$ and $\omega_0=1/\sqrt{\mathrm{LC}}$. 
Then, Fig. \ref{fig:Single-LC-resonator} leads to  
\begin{equation}
    \frac{da_{\pm}}{dt}=\pm j\omega_0a_{\pm} \label{eq:a}\,.
\end{equation}
$a_{+}$ and $a_{-}$ are called the positive and negative frequency components of the mode amplitude, and always satisfy 
\begin{equation}
    a_{-}=\left(a_{+}\right)^{*},
\end{equation}
where "*" denotes complex conjugate.
The amplitudes $a_\pm$ in (\ref{eq:12}) are defined such that both terms in the parenthesis have the same unit, and the total energy in the resonator $W$ is 
\begin{equation}
    W= a_+a_-= \left|a_+\right|^2.\label{eq:W_a}
\end{equation}
The two equations in (\ref{eq:a}) are decoupled and only one of them needs to be solved. Note that solving (\ref{eq:a}) has the same complexity as solving (\ref{eq:3}) or (\ref{eq:4}).
It requires solving first order differential equations in the complex numbers space instead of solving second order differential equations in the real numbers space. However, complex mode amplitudes are more suitable of studying the energy of a system in time domain. In fact, elevating them to operators and applying the scaling factor $1/\sqrt{\hbar \omega_0}$ leads to the creation and annihilation operators in circuit/cavity quantum electrodynamics (see appendix \ref{sec:quantized fields in resonators}.) In other words, 
\begin{equation}
    a_+ \Longleftrightarrow \sqrt{\hbar \omega_0}\hat{a}.\label{eq:classic_quantum}
\end{equation}
The quantum operators are defined to be dimensionless so that the energy of the system is $W=\hbar \omega_0 \hat{a}^{\dagger}\hat{a}$, instead of \eqref{eq:W_a}. 

If the resonator also includes lossy elements $\mathrm{R}$ and $\mathrm{G}$, as shown in Fig. \ref{fig:Single-LC-resonator}, it is straightforward to show 
\begin{equation}
   \frac{da_{\pm}}{dt}=\pm j\omega_{0}a_{\pm}-\frac{1}{2}\left(\frac{\mathrm{G}}{\mathrm{C}}\pm\frac{\mathrm{R}}{\mathrm{L}}\right)a_{+}-\frac{1}{2}\left(\frac{\mathrm{G}}{\mathrm{C}}\mp\frac{\mathrm{R}}{\mathrm{L}}\right)a_{-}. \label{eq:da_exact}
\end{equation}

\begin{itemize}
    \item if $\mathrm{RC=GL}$, the two equations in \eqref{eq:da_exact} are decoupled, 
    
\begin{equation}
   \frac{da_{\pm}}{dt}=\pm j\omega_{0}a_{\pm}-\frac{1}{2}\left(\frac{\mathrm{G}}{\mathrm{C}}+\frac{\mathrm{R}}{\mathrm{L}}\right)a_{\pm}.
\end{equation}
That is, the energy decays exponentially in time, without any oscillation,
\begin{equation}
    \frac{dW}{dt}=-\kappa W,
\end{equation}
in which $\kappa=\left(\frac{\mathrm{G}}{\mathrm{C}}+\frac{\mathrm{R}}{\mathrm{L}}\right)$ is the energy decay rate. 
\item if $\mathrm{RC\neq GL}$, the two equations in \eqref{eq:da_exact} remain coupled, and the energy decays as \begin{equation}
    \frac{dW}{dt}=-\kappa W-\frac{1}{2}\left(\mathrm{\frac{G}{C}-\frac{R}{L}}\right)\left(a_{+}a_{+}+a_{-}a_{-}\right) \label{eq:dW_exact}.
\end{equation}
In other words, the energy has fast oscillations in time, but its moving average decays exponentially. Note that the amplitude of the oscillation decreases as $\mathrm{RC}$ approaches $\mathrm{GL}$. 
\end{itemize}

It is common to drop one of the terms in \eqref{eq:da_exact} and obtain a decoupled set of equations, a.k.a. Rotating Wave Approximation (RWA),  as
\begin{equation}
    \frac{d a_\pm}{dt } =\pm j\omega_0 a_\pm -\frac{\kappa}{2} a_\pm. \label{eq:da}
\end{equation}
This is equivalent to ignoring the last term in \eqref{eq:dW_exact}.

 $\mathrm{R}$ and $\mathrm{G}$ in Fig. \ref{fig:Single-LC-resonator} indicate the total energy loss experienced by the resonator, and can include couplings to the environment (e.g., a transmission line). Because the focus of this document is on the couplings, let us assume the resonator has zero intrinsic loss for the remainder of the discussions.

\section{A note on distributed resonators}

\begin{figure}[h]
    \centering
    \subfloat[]{\includegraphics[width=0.4\textwidth]{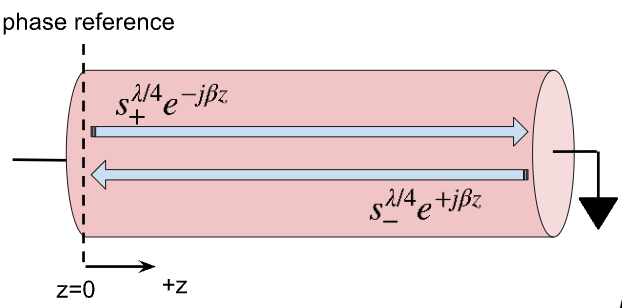}}
    
    \subfloat[]{\includegraphics[width=0.4\textwidth]{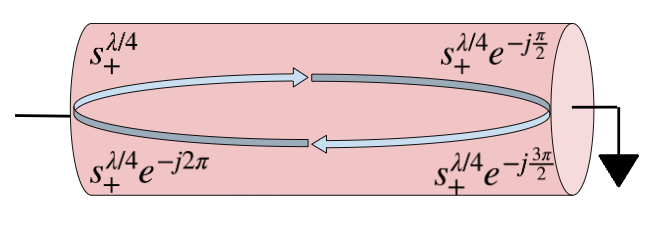}}
    \caption{A quarter-wave resonator described by a) two travelling power waves in opposite directions, and b) a circulating power wave. Both pictures lead to the same conclusions.}
    \label{fig:circulating_s+}
\end{figure}

The resonance mode amplitudes $a_{\pm}$ in \eqref{eq:a} are defined based on the lumped element model of the resonator. In general, the mode amplitudes of any electromagnetic resonator can be obtained from its fields (e.g., see Appendix \ref{sec:quantized fields in resonators}). However, distributed resonators using uniform transmission lines are very common in superconducting devices. This is partly because their fields can be confined to a local region, and their design (e.g., coupling to a transmission line) are straightforward. In frequency domain (steady state), the resonant mode in a distributed resonator is formed by the interference of two power waves traveling in opposite directions, $s^{\mathrm{res}}_\pm$, defined as 
\begin{equation}
        s^{\mathrm{\mathrm{res}}}_{\pm}e^{\mp j\beta z}=\frac{1}{\sqrt{2}}\sqrt{\oint dA \, \hat{z}\cdot\left(\vec{E}_{\mathrm{t}}\times\vec{H}_{\mathrm{t}}^{*}\right)_\pm}\,\,e^{\mp j\beta z},
\end{equation} 
where the subscript "t" denotes the transverse fields to the direction of propagation, the integration is over the cross section of the transmission line, $\pm \mathrm{z}$ is the propagation direction, and $\beta$ is the propagation constant. The two power waves, in most resonators, are not independent. For example, in a quarter-wave resonator, the two power waves are equal for the phase reference chosen on the resonator's open end. That is, $s_-^{\lambda/4}=s_+^{\lambda/4}$ in Fig. \ref{fig:circulating_s+}(a). In some resonators, e.g. ring resonators, the phases of the two modes can remain uncorrelated, and therefore degenerate modes can exist.

The power traveling towards +z direction through the cross section of the resonator is 
\begin{equation}
    P_+= s_+^{\mathrm{res}}\left(s_+^{\mathrm{res}}\right)^*=\left|s_+^{\mathrm{res}}\right|^2,
\end{equation}
using Poynting's theorem. In order to find the relation between the mode amplitude $a_+$ and the power wave $s_+^{\mathrm{res}}$, consider a quarter-wave resonator as shown in Fig. \ref{fig:circulating_s+}. The resonator's power wave is denoted by $s_\pm^{\lambda/4}$ for clarity.

The travel time of (the wave front of) the power waves between the two boundaries of the resonator is $t=1/(4f_0)$, where $f_0$ is the resonance frequency. Therefore, the total energy required to populate the resonator with both power waves is 
\begin{equation}
    W=2\int_0^{\frac{1}{4f_0}}dt\left|s^{\lambda/4}_+\right|^2=\frac{\left|s^{\lambda/4}_+\right|^2}{2f_0},\label{eq:W_s_res}
\end{equation}
in which $\left|s_+^{\lambda/4}\right|=\left|s_-^{\lambda/4}\right|$ is used. The second equality in (\ref{eq:W_s_res}) is with the assumption that the resonator is lossless and therefore $\left|s^{\lambda/4}_+\right|$ is independent of z. Using \eqref{eq:W_a} and \eqref{eq:W_s_res},
\begin{equation}
    \left|a_+\right|=\frac{\left|s_+^{\lambda/4}\right|}{\sqrt{2 f_0}}, \label{eq:powerwave_quarterwave}
\end{equation}
for a properly chosen phase reference point in Fig. \ref{fig:circulating_s+}. 

Alternatively, we can consider a circulating power wave $s_+^{\lambda/4}e^{-j\beta r}$ inside the resonator  where r is the travel direction and is +z(-z) in the first(second) half of circulation path. The reflection from the short end of the resonator adds an additional $\pi$ phase shift to the power wave. This is clarified in Fig. \ref{fig:circulating_s+}(b).

Similarly, the power wave in a half-wave resonator $s_+^{\lambda/2}$ is related to $a_+$ as 
\begin{equation}
    \left|a_+\right|=\frac{2\left|s_+^{\lambda/2}\right|}{\sqrt{2f_0}}. \label{eq:powerwave_halfwave}
\end{equation}

Relations \eqref{eq:powerwave_quarterwave} and \eqref{eq:powerwave_halfwave} are very useful in analyzing systems where the coupling between a distributed resonator and a transmission line is mediated by a microwave coupler. This will be reviewed in a later section. 

For the sake of completeness, the relations between the power wave and the voltage and current waves in transmission line theory are \cite{collin2007foundations}
\begin{equation}
    v_+=\sqrt{2Z_{\mathrm{w}}}s_+^{\mathrm{res}}; \quad
    i_+=\sqrt{\frac{2}{Z_{\mathrm{w}}}}s_+^{\mathrm{res}}, \label{eq:power-to-tl}
\end{equation}
in which the wave impedance of the mode, $Z_{\mathrm{w}}=\left|\vec{E}_{\mathrm{t}}/\vec{H}_{\mathrm{t}}\right|$, is used. However, the impedance in \eqref{eq:power-to-tl} is an arbitrary choice and, in general, current (voltage) amplitude in transmission line theory is not always uniquely defined. In special cases, such as two-conductor TEM transmission lines, the common definitions of voltage and current are applicable, which also coincide with \eqref{eq:power-to-tl}.

\section{A resonator coupled to lossy environment}
\leveldown{Singly loaded resonator}

\begin{figure}[h]
    \centering
    \subfloat[Single-ended transmission line coupled to a resonator.]{\includegraphics[clip,width=0.4\textwidth]{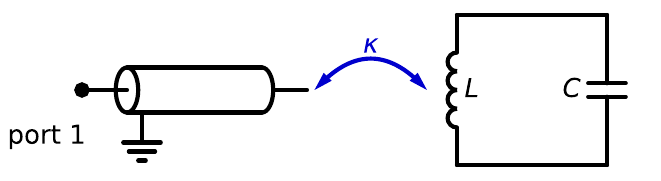}}
    
    \subfloat[Reflection phase of the transmission line versus frequency.]{\includegraphics[clip,width=0.4\textwidth]{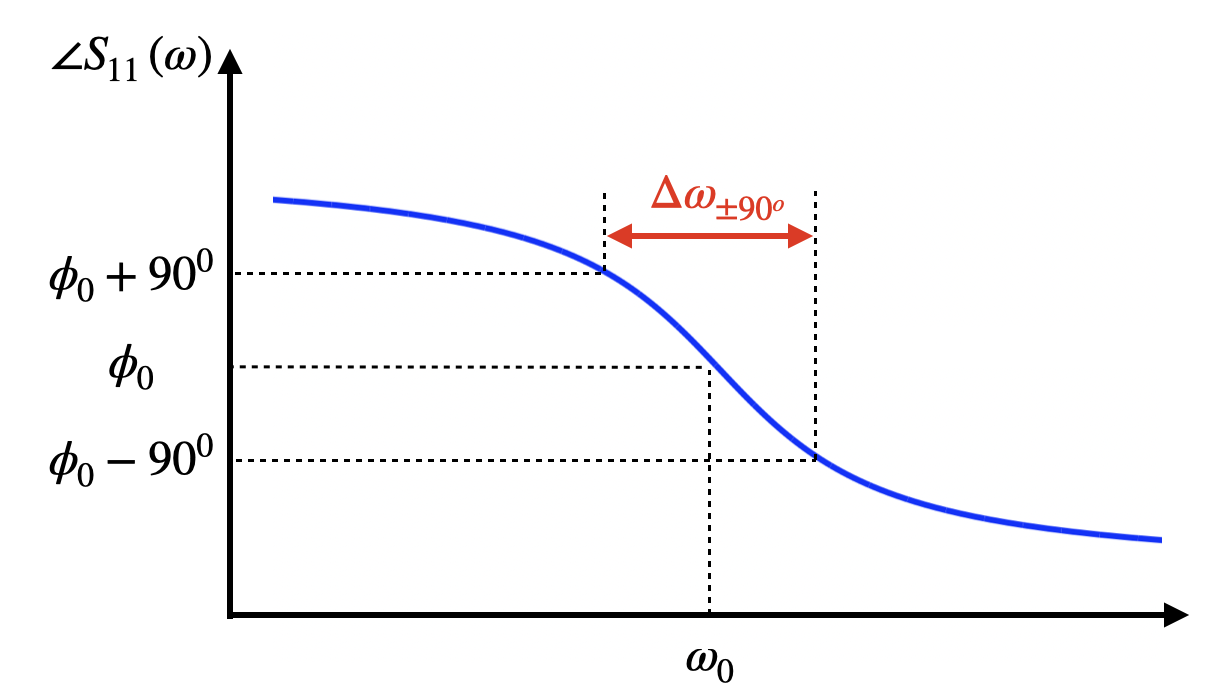}}
    \caption{Singly loaded resonator.}
    \label{fig:SLR}
\end{figure}

Consider a transmission line terminated to a resonator via the coupling $\kappa$. This coupling can be mediated via the overlapping electric and magnetic fields of the resonator and transmission line. If there is an incoming power wave on the transmission line $s_+$ bringing energy to the resonator and a reflected power wave $s_-$ carrying energy away from the resonator, then
\begin{subequations}
    \begin{align}
        \frac{d a_+}{d t} & = j\omega_0 a_+ + \frac{\sqrt{\kappa_+}}{2}s_+ + e^{j\phi_0} \frac{\sqrt{\kappa_-}}{2}s_-,\label{eq:lossy_a_plus} \\
        \frac{d a_-}{d t} & = - j\omega_0 a_- + \frac{\sqrt{\kappa_+}}{2}s_+^* + e^{-j\phi_0} \frac{\sqrt{\kappa_-}}{2}s_-^*.\label{eq:lossy_a_minus} 
    \end{align}
\end{subequations}
where $\phi_0$ is determined by the considered phase reference point in the transmission line as shown in Fig. \ref{fig:SLR}. Later, it becomes apparent that $\phi_0$ is the reflection phase on the transmission line in the absence  of the resonator.

We have asserted $a_\pm = a_\mp^*$ remains true. The power waves are defined such that the incident power on the resonator is $P_{\text{inc}}=\left|s_+\right|^2$ and the reflected power from the resonator is $P_{\text{ref}}=\left|s_-\right|^2$. Note that the standard definition of power waves in electrical engineering does not include the $\omega_0$ normalization coefficient. It is added to simplify the formulation. The time convention for the power waves is chosen to match the frequency sign of $a_+$. The incident and reflected power wave couplings to $a_+$ are $\sqrt{\kappa_\pm}$ respectively. The couplings between the power waves and $a_-$ are neglected because their frequencies have opposite signs (i.e., they are too far away from each other in frequency space). This is another approximation, besides RWA, that is often used in studying a coupled resonator-transmission line. 

Under time reversal, we have $a_\pm \to a_\mp$ and $s_\pm \to s_\mp^*$. If we demand time reversal symmetry, then $\kappa_+ = \kappa_- = \kappa$.

The net power delivered to the resonator is given by
\begin{equation}
    \frac{dW}{dt} = \frac{d(a_+a_-)}{dt} = |s_+|^2 - |s_-|^2. \label{eq:power}
\end{equation}
If we take $s_-$ to be an output of incoming power and system dynamics, then the following linear combination uniquely satisfies \eqref{eq:lossy_a_plus}, \eqref{eq:lossy_a_minus} and \eqref{eq:power}:
\begin{equation}
    s_- = e^{-j\phi_0}\left(s_+-\sqrt{\kappa} a_+ \right), \label{eq:io}
\end{equation}
often referred to as the input-output relation. If substituted back into \eqref{eq:lossy_a_plus}, we get the familiar form of the Langevin equation:
\begin{equation}
    \frac{d a_+}{d t} =  j\omega_0 a_+ -\frac{\kappa}{2} a_++\sqrt{\kappa} s_+.\label{eq:lossy_dadt} 
\end{equation}
Note that although $a_-=a_+^*$ is always true, $s_-$ is not necessarily equal to $s_+^*$. Also, in the absence of the resonator, i.e. if $\kappa$ in \eqref{eq:io} becomes zero, 
\begin{equation}
    s_-=\left.s_+e^{-j\phi_0}\right|_{\kappa\rightarrow0}.
\end{equation}
which clarifies the definition of the phase reference point in Fig. \ref{fig:SLR}. For example, if the transmission line is terminated to a short, $\phi_0=\pi$ since the reflection coefficient from a short boundary is $-1$. 
If  $s_+$ is harmonic with the frequency of $\omega$, the steady-state response of the resonator is obtained by taking the Fourier transform of (\ref{eq:lossy_dadt}),
\begin{equation}
    a_+(\omega)=\frac{\sqrt{\kappa} s_+(\omega)}{j\left(\omega-\omega_0\right)+\kappa/2}.\label{eq:lossy_w}
\end{equation}
It can be shown that $\kappa$ is also the resonator`s bandwidth used in calculating the resonance quality factor, $Q=\frac{\omega_0}{\kappa}$.

If the resonator is fed at its resonance frequency, the resonator's amplitude in steady state is 
\begin{equation}
    a_+(\omega_0)=\frac{2}{\sqrt{\kappa}}s_+(\omega_0).
\end{equation}
As expected, the resonator's amplitude increases by decreasing $\kappa$. It also increases the time needed to energize the resonator to a target amount.

The reflection coefficient of the resonator in the steady state can be found using (\ref{eq:io}) and (\ref{eq:lossy_w})   as 
\begin{equation}
    S_{11}\left(\omega\right)=\frac{s_-\left(\omega\right)}{s_+\left(\omega\right)}=e^{-j\phi_0}\frac{j\left(\omega-\omega_0\right)-\kappa/2}{j\left(\omega-\omega_0\right)+\kappa/2}. \label{eq:S11_singly_loaded}
\end{equation}
As expected, the reflection amplitude is unity in steady-state. Also, derivative of the reflection phase is 
\begin{equation}
   \frac{\partial \angle{S_{11}\left(\omega\right)}}{\partial \omega}=-\frac{4}{\kappa}\left(1+\left(\frac{\omega-\omega_0}{\kappa/2}\right)^2\right)^{-1}.\label{eq:reflection_derivative}
\end{equation}
Equation (\ref{eq:reflection_derivative}) implies three important conclusions: 
\begin{enumerate}
    \item there is an inflection point at $\omega=\omega_0$,
\begin{equation}
    \left.\frac{\partial^2 \angle S_{11}(\omega)}{\partial \omega^2}\right|_{\omega=\omega_0}=0.
\end{equation}
    \item $\kappa$ can be obtained from 
        \begin{equation}
            \kappa=-4\Bigg/\left.\frac{\partial \angle{S_{11}\left(\omega\right)}}{\partial \omega}\right|_{\omega=\omega_0}
        \end{equation}
    \item  $\omega-\omega_0=\pm \frac{\kappa}{2}$ leads to $\angle{S_{11}}\left(\omega\right)=-\phi_0\pm \frac{\pi}{2}$, which means $\kappa$ and the loaded quality factor of the resonator can be extracted from the phase response of the transmission line in frequency domain as  
        \begin{equation}
            Q_{\mathrm{e}}^{\mathrm{SL}}=\frac{\omega_0}{\kappa}=\frac{\omega_0}{\Delta \omega_{\pm 90^o}}. \label{eq:Q_SE}
        \end{equation}
        $Q_{\mathrm{e}}^{\mathrm{SL}}$ is the quality factor of the singly loaded resonator, and $\Delta\omega_{\pm 90^o}$ is the $\pm 90^o$ phase change around the resonance frequency, as illustrated in Fig. \ref{fig:SLR}. 
\end{enumerate}
Equation \eqref{eq:S11_singly_loaded} can also be obtained by finding the impedance of the resonator in frequency domain and using the approximation $(\omega^2-\omega_0^2)/\omega \simeq 2 \Delta \omega$. [\cite{hong2004microstrip}, p. 260]. This is equivalent of RWA used in \eqref{eq:da}.

\levelstay{A resonator loaded with an N-port network}

\begin{figure}[t]
    \centering
    \includegraphics[width=0.3\textwidth]{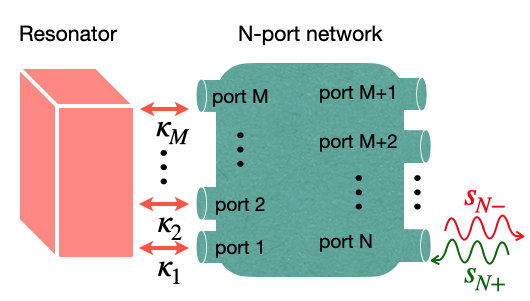} 
    \caption{An N-port network coupled to a resonator through M couplings.}
    \label{fig:N-port}
\end{figure}

Consider a resonator coupled to a lossless reciprocal N-port network through M(<=N) ports as shown in Fig. \ref{fig:N-port}. The scattering coefficients that are coupled (C) and independent (I) to a resonance and follow:
\begin{equation}
  \begin{pmatrix} \mathbf{s}_{\mathrm{C-}} \\ \mathbf{s}_{\mathrm{I-}} \end{pmatrix} =
     \mathbf{S} \begin{pmatrix} \mathbf{s}_{\mathrm{C+}} \\ \mathbf{s}_{\mathrm{I+}} \end{pmatrix},
\end{equation}
where 
\begin{equation}
    \mathbf{S} =   \begin{pmatrix} \mathbf{s}_{\mathrm{CC}} & \mathbf{\mathbf{s}_{\mathrm{CI}}} \\ \mathbf{s}_{\mathrm{IC}} & \mathbf{s}_{\mathrm{II}} \end{pmatrix} \label{eq:S_matrix}
\end{equation}
is an N$\times$N matrix in which $\mathbf{s_{\mathrm{CC}}}$ and $\mathbf{s_{\mathrm{II}}}$ are M$\times$M and (N$-$M)$\times$(N$-$M) matrices, respectively. 
The scattering matrix of the N-port network also satisfies the unitary condition  
\begin{equation}
    \mathbf{SS^{\dagger}=I}_{\mathrm{N\times N}},
\end{equation}
where superscript $\dagger$ denotes transposed complex conjugate of the matrix. 
Similar to the singly loaded resonator, one can start with 
\begin{equation}
   \frac{da_{+}}{dt}=j\omega_{0}a_{+}+\frac{\sqrt{\boldsymbol{\kappa}}^{\mathrm{t}}}{2}\left(\mathbf{s}_{\mathrm{C-}}+e^{j\boldsymbol{\phi}_0}\mathbf{s}_{\mathrm{C+}}\right), \label{eq:da_Nport}
\end{equation}
and look for an input-output relation that satisfies energy conservation. In \eqref{eq:da_Nport}, 
\begin{equation}
    \boldsymbol{\phi}_0=\begin{pmatrix} \phi_0^1 & 0 & \cdots & 0 \\ 0 & \phi_0^2  & \cdots & 0 \\ \vdots & & & \vdots \\ 0 & 0 & \cdots & \phi_0^{\mathrm{M}} \end{pmatrix},
\end{equation}
\begin{equation}
    \sqrt{\boldsymbol{\kappa}}=\begin{pmatrix} \sqrt{\kappa_1} \\ \sqrt{\kappa_2} \\ \vdots \\ \sqrt{\kappa_{\mathrm{M}}} \end{pmatrix}
\end{equation}
are the reflection phase and coupling matrices, respectively. Note that the incoming and outgoing power waves are defined with reference to the N port network, hence the difference between \eqref{eq:da_Nport} and \eqref{eq:lossy_a_plus}.

The energy conservation imposes
\begin{equation}
    \frac{d\left( a_+a_-\right)}{dt}=\mathbf{s_{\mathrm{C}-}^{\dagger}s_{\mathrm{C}-}-s_{\mathrm{C}+}^{\dagger}s_{\mathrm{C}+}}. \label{eq:energy_matrix}
\end{equation}
It can be shown that the unique non-trivial solution of \eqref{eq:energy_matrix} is
\begin{equation}
     \mathbf{s}_{\mathrm{C+}} = e^{-j\boldsymbol{\phi}_0}\left( \mathbf{s}_{\mathrm{C-}}- \sqrt{\boldsymbol{\kappa}} a_+\right).
\end{equation}
Therefore, the equation of motion of the resonator can be expressed as  
\begin{equation}
  \frac{da_{+}}{dt}=j\omega_{0}a_{+}+\sqrt{\boldsymbol{\kappa}}^{\mathrm{t}}\left(I-e^{-j\boldsymbol{\phi}_{0}}\mathbf{s}_{\mathrm{CC}}\right)^{-1}\left(\mathbf{\mathbf{s}_{\mathrm{CI}}}\mathbf{s}_{\mathrm{I+}}-\frac{1}{2}\sqrt{\boldsymbol{\kappa}} a_+\right) \label{eq:EOM_Nport}
\end{equation}
% \begin{equation*}
%     \qquad\qquad-\frac{1}{2}\sqrt{\kappa}^{t}\left(I-e^{-j\phi_{0}}\mathbf{s_{\mathrm{CC}}}\right)^{-1}\sqrt{\kappa}a_{+}.
% \end{equation*}
The scattering relation of the reduced N-port network is 
\begin{equation}
    \mathbf{s}_{\mathrm{I-}} = \left(\mathbf{s}_{\mathrm{II}} + e^{-j\boldsymbol{\phi}_{0}}\mathbf{s}_{\mathrm{IC}}\left(I-e^{-j\boldsymbol{\phi}_{0}}\mathbf{s}_{\mathrm{CC}}\right)^{-1}\mathbf{s}_{\mathrm{CI}}\right)\mathbf{s}_{\mathrm{I+}}\label{eq:reduced_S}
\end{equation}
\begin{equation*}
    -\left(\mathbf{s}_{\mathrm{IC}} + e^{-j\boldsymbol{\phi}_0}\mathbf{s}_{\mathrm{IC}}\left(I-e^{-j\boldsymbol{\phi}_{0}}\mathbf{s}_{\mathrm{CC}}\right)^{-1}\mathbf{s}_{\mathrm{CC}}\right)e^{-j\boldsymbol{\phi}_{0}}\sqrt{\boldsymbol{\kappa}}a_{+}. 
\end{equation*}

The relations of a singly loaded resonator, discussed in previous section, can be easily obtained by considering a two port network coupled to a resonator in  \eqref{eq:EOM_Nport} and \eqref{eq:reduced_S}. In the following, we consider two more examples: a doubly loaded resonator and a resonator coupled to a transmissive line.  

\levelstay{Doubly loaded resonator}
\begin{figure}[t]
    \centering
    \includegraphics[width=0.5\textwidth]{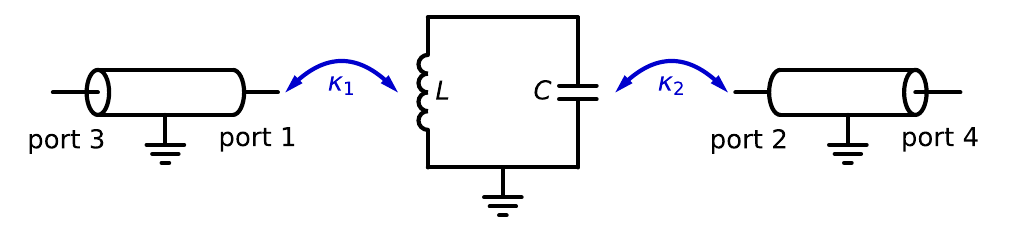}       
    \caption{Doubly loaded resonator.}
    \label{fig:DLR}
\end{figure}

Consider a resonator coupled to two transmission lines as shown in Fig. \ref{fig:DLR}. This system is also known as a doubly loaded resonator. The two transmission lines form a four port network with the scattering matrix
\begin{equation}
    \mathbf{S}=\begin{pmatrix} 0 & 0 & e^{-j\theta_1} & 0 \\ 0 & 0 & 0 & e^{-j\theta_2} \\e^{-j\theta_1} & 0 & 0 & 0 \\ 0 & e^{-j\theta_2} & 0 & 0
    \end{pmatrix},
\end{equation}
where $\theta_1$ and $\theta_2$ are electrical lengths of the two transmission lines. 
Based on \eqref{eq:S_matrix}, 
\begin{equation*}
      \mathbf{s}_{\mathrm{CC}}=\mathbf{s}_{\mathrm{II}}=\mathbf{0},
\end{equation*}
\begin{equation}
      \mathbf{s}_{\mathrm{IC}}=\mathbf{s}_{\mathrm{CI}}=\begin{pmatrix} e^{-j\theta_1} & 0  \\ 0 & e^{-j\theta_2}
      \end{pmatrix}.
\end{equation}
The coupling matrix is 
\begin{equation}
   \sqrt{\boldsymbol{\kappa}}=\begin{pmatrix}  \sqrt{\kappa_1} \\ \sqrt{\kappa_2}
      \end{pmatrix}. 
\end{equation}
Therefore, \eqref{eq:EOM_Nport} and \eqref{eq:reduced_S} lead to
\begin{equation}
      \frac{d a_+}{d t} = j\omega_0 a_+ -\frac{\kappa_1+\kappa_2}{2} a_++ \sqrt{\kappa_1}e^{-j\theta_1}s_{3+}+\sqrt{\kappa_2}e^{-j\theta_2} s_{4+},\label{eq:doubly_loaded_res} 
\end{equation}
\begin{equation}
    \left(\begin{array}{c}
s_{3-}\\
s_{4-}
\end{array}\right)=\left(\begin{array}{cc}
e^{-j2\theta_{1}} & 0\\
0 & e^{-j2\theta_{2}}
\end{array}\right)\left(\begin{array}{c}
s_{3+}\\
s_{4+}
\end{array}\right)-\left(\begin{array}{c}
\sqrt{\kappa_{1}}e^{-j\theta_{1}}\\
\sqrt{\kappa_{2}}e^{-j\theta_{2}}
\end{array}\right)a_{+},\label{eq:reflections_doubly_loaded}
\end{equation} 
where $\boldsymbol{\phi}_0=\mathbf{0}$ is used.  

\begin{figure}[t]
    \centering
    \includegraphics[width=0.4\textwidth]{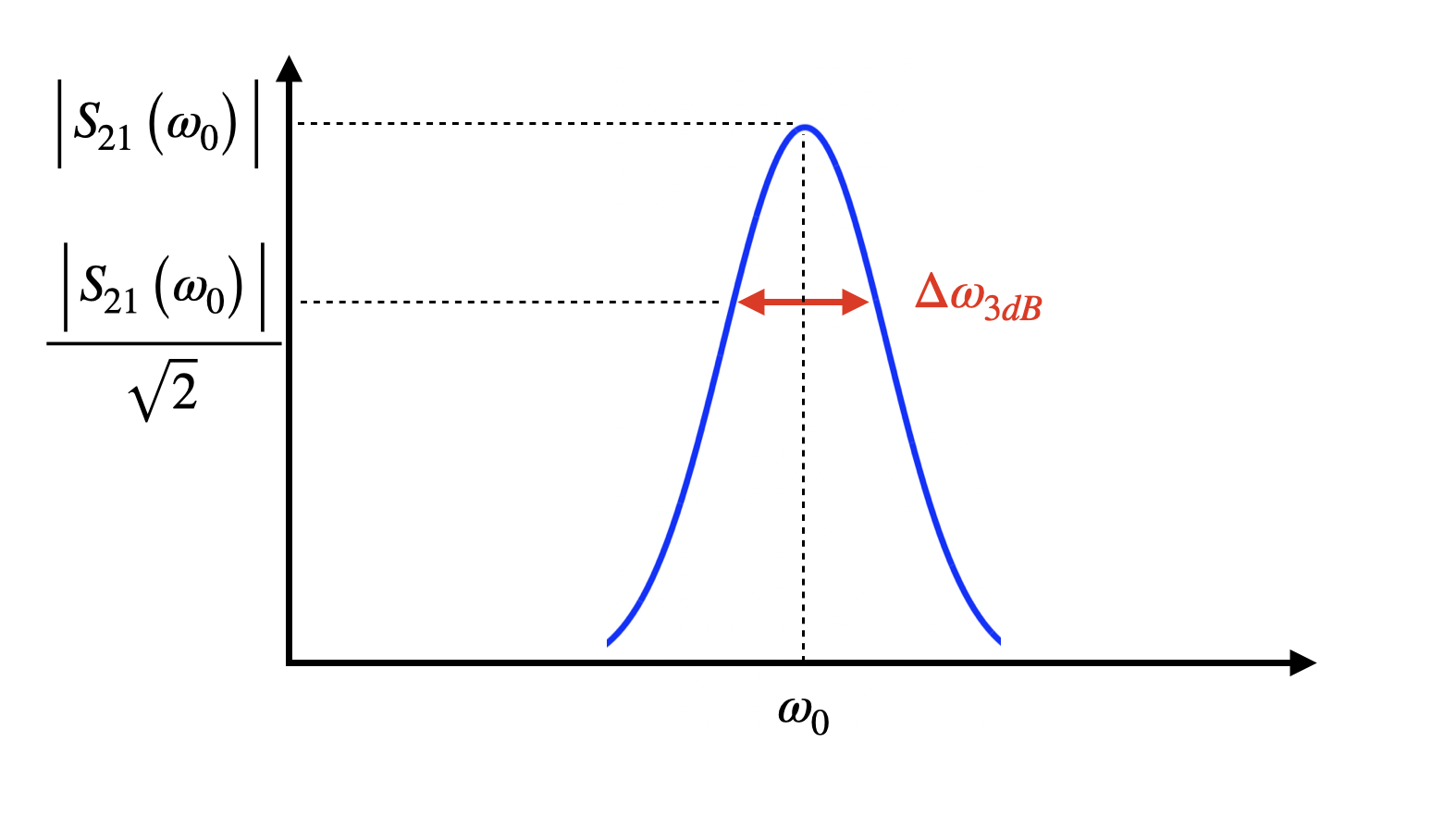} 
    \caption{Transmission through a doubly loaded resonator.}
    \label{fig:DLR_S21}
\end{figure}

Using (\ref{eq:doubly_loaded_res}) at steady-state and \eqref{eq:reflections_doubly_loaded}, the transmission through the system is 
\begin{equation}
    S_{43}=\frac{s_{4-}}{s_{3+}}=-\frac{\sqrt{\kappa_{1}\kappa_{2}}e^{-j\left(\theta_1+\theta_2\right)}}{j\left(\omega-\omega_{0}\right)+\left(\kappa_{1}+\kappa_{2}\right)/2}. \label{eq:transmission_doubly_loaded}
\end{equation}
 The maximum transmission occurs at $\omega=\omega_0$. Also,  $\left(\omega-\omega_0\right)=\pm \frac{\kappa_1+\kappa_2}{2}$ leads to $\left|S_{43}\right|=\frac{1}{\sqrt{2}}\left|S_{43}\right|_{max}$. In other words, $\kappa_1+\kappa_2$ and the loaded quality factor of the resonator can be extracted from   
        \begin{equation}
           \kappa_1+\kappa_2=\frac{\omega_0}{ Q_e^{DL}}=\Delta \omega_{3dB}, \label{eq:Q_SE}
        \end{equation}
where $Q_e^{DL}$ is the quality factor of the doubly loaded resonator, and $\omega_{3dB}$ is illustrated in Fig. \ref{fig:DLR_S21}.
Also, if $\kappa_1=\kappa_2$, the transmission through the resonator is always unity at the resonance frequency. This is independent of $\kappa$, which is very important. 

\levelstay{Resonator coupled to a transmissive path}

\begin{figure}[t]
    \centering
    \subfloat[A resonator weakly coupled to a transmissive line.]{\includegraphics[clip,width=0.4\textwidth]{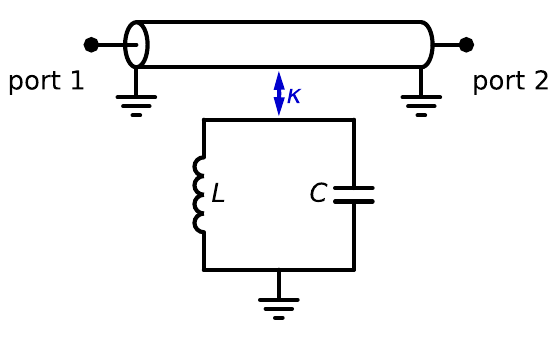}} 
    
    \subfloat[Equivalent to (a), represented by a microwave. T-junction.]{\includegraphics[clip,width=0.4\textwidth]{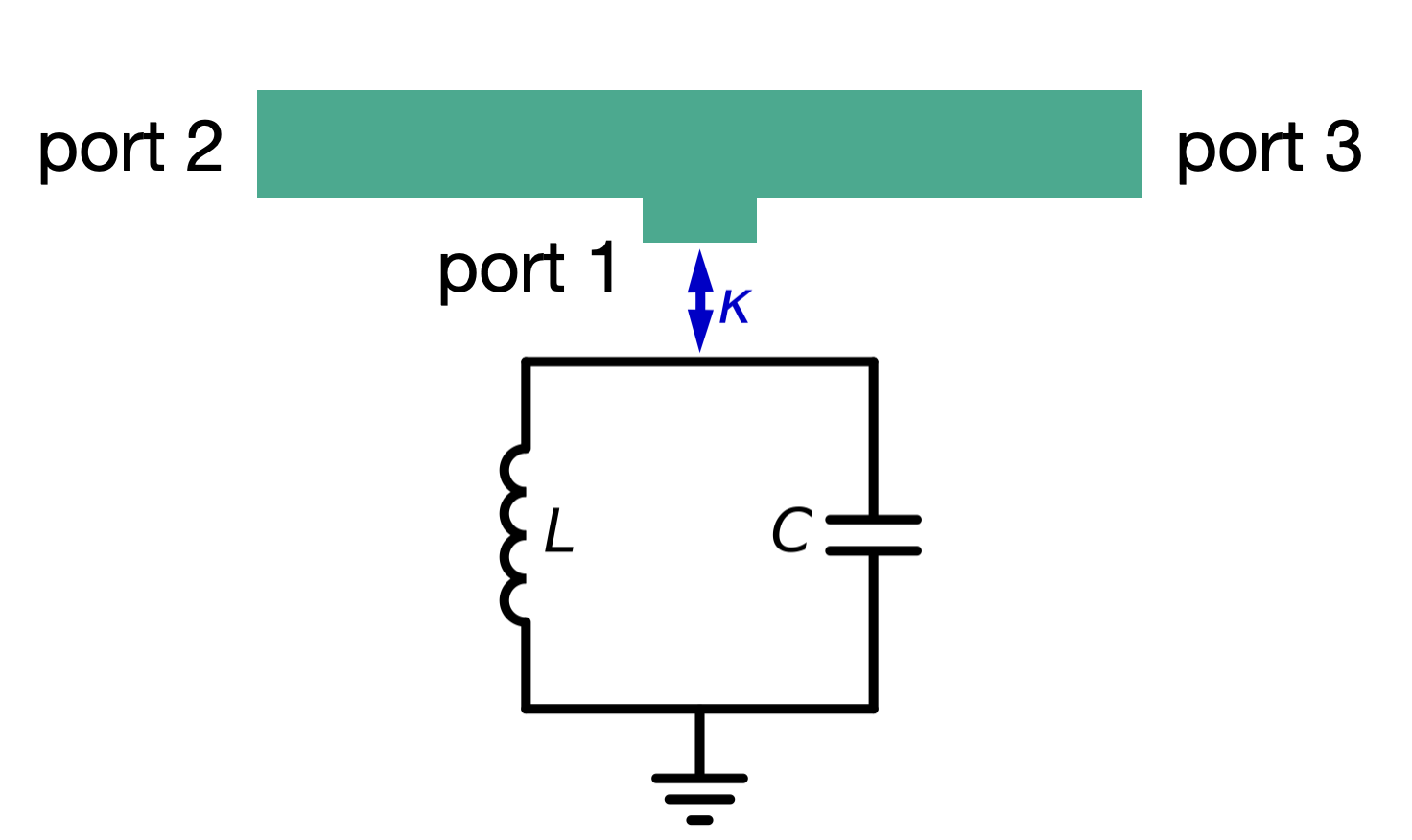}}
    \caption{A resonator, side-coupled to a transmission line.}
    \label{fig:parallel_coupled}
\end{figure}

Another common geometry in superconducting devices is a resonator that is weakly coupled to a transmission line as shown in Fig. \ref{fig:parallel_coupled}. This geometry can be represented by a T-junction in which port 1 is coupled to the resonator. In the absence of the resonator, port 1 is open, leaving a reflection-free path between ports 2 and 3. The scattering matrix of a symmetrical T junction is 
\begin{equation}
    \mathbf{S}=\begin{pmatrix} -\frac{1}{3} &\frac{2}{3} &\frac{2}{3} \\ \frac{2}{3}&-\frac{1}{3}&\frac{2}{3} \\ \frac{2}{3}&\frac{2}{3}&-\frac{1}{3}
    \end{pmatrix}, \label{eq:Tjunction}
\end{equation}
assuming the T-junction's dimensions are much smaller than the wavelength. 
Therefore, 
\begin{equation}
      \mathbf{s}_{\mathrm{CC}}=-\frac{1}{3},\quad \mathbf{s}_{\mathrm{IC}}=\mathbf{\mathbf{s_{\mathrm{CI}}}}^t=\begin{pmatrix} \frac{2}{3} \\ \frac{2}{3} 
      \end{pmatrix}, \quad  \mathbf{s_{\mathrm{II}}}=\begin{pmatrix} -\frac{1}{3} & \frac{2}{3}  \\ \frac{2}{3} & -\frac{1}{3}
      \end{pmatrix}.
\end{equation}
Then \eqref{eq:EOM_Nport} and \eqref{eq:reduced_S} give
\begin{equation}
    \frac{da_{+}}{dt}=j\omega_{0}a_{+}-\frac{\kappa}{4} a_{+}+\frac{\sqrt{\kappa}}{2}\left(s_{2+}+s_{3+}\right),\label{eq:EOM_Tee}
\end{equation}
\begin{equation}
    \begin{pmatrix} s_{2-} \\ s_{3-} \end{pmatrix}=\left(\begin{array}{cc}
0 & 1\\
1 & 0
\end{array}\right)\begin{pmatrix} s_{2+} \\ s_{3+} \end{pmatrix}- \frac{\sqrt{\kappa}}{2} \left(\begin{array}{c}
1\\
1
\end{array}\right)a_{+}.
\end{equation}
Transmission through the system in steady-state is
\begin{equation}
    S_{32}=\frac{s_{3-}}{s_{2+}}=\frac{j\left(\omega-\omega_{0}\right)}{j\left(\omega-\omega_{0}\right)+\kappa/4},
\end{equation}
which is maximally disturbed (becomes zero) at the resonance frequency. 
If the resonator has intrinsic loss, the non-zero transmission at resonance can be used to extract the intrinsic loss. 
From \eqref{eq:EOM_Tee}, the energy decay rate of the resonator is $\kappa/2$. Similar to doubly-loaded resonator, $\kappa$ can be extracted from  the transmission spectrum as 
\begin{equation}
    \frac{\kappa}{2}=\Delta\omega_{3dB},
\end{equation}
where $S_{32}\left(\omega_{3dB}\right)=\frac{1}{\sqrt{2}}.$

\section{Resonator-resonator coupling}

Coupled resonators are best described by considering them as a unified multi-mode resonator and extracting its eigenmodes.
However, when the resonators are weakly coupled, it is also desirable to represent the coupled resonators with their individual isolated modes and defining coupling coefficients among them \cite{cohn1968microwave,van1982weakly,zaki1987coupling}. The unit-less coupling coefficient between two resonators is defined as \cite{hong2004microstrip,hong2000couplings}
\begin{equation}
  \zeta
  =\frac{\int dv \, \varepsilon\Vec{E_{1}} \cdot \Vec{E_2}}{\sqrt{\int dv \, \varepsilon\abs{E_1}^2  \times\int dv \, \varepsilon \abs{E_2}^2}} + \frac{\int dv \, \mu\Vec{H_1} \cdot \Vec{H_2}}{\sqrt{\int dv \, \mu\abs{H_1}^2 \times \int dv \,\mu \abs{H_2}^2}}
  \label{eq:cc}
\end{equation}
where $\Vec{E}_{1,2}$ and $\Vec{H}_{1,2}$ are the electric and magnetic fields intensities at bare resonance frequencies of the resonators, $\mu$ is the permeability, $\varepsilon$ is the permittivity and the integrals are over the entire volume.
The fields subscript 1(2) refer to the fields of the resonator 1(2) after replacing the resonator 2(1) with the ambient medium of resonator 1(2).

Equation \eqref{eq:cc} defines the coupling as the sum of the ratios of the coupled electric(magnetic) energy to the geometric mean of the stored electric(magnetic) energies in both resonators. The reason for choosing this definition will become apparent soon. 
Calculating \eqref{eq:cc} is cumbersome since the fields of the resonators can be at different frequencies, and multiple geometries need to be solved. There are also subtleties in defining the bare modes of the resonators or in the presence of surface currents  \cite{elnaggar2015energy}. Usually, alternative approaches are used. In the following, the circuit equivalent of \eqref{eq:cc} is extracted for two  coupled resonators using Kirchhoff's laws.
 
\begin{figure}[t!]
    \centering
    \includegraphics[width=0.4\textwidth]{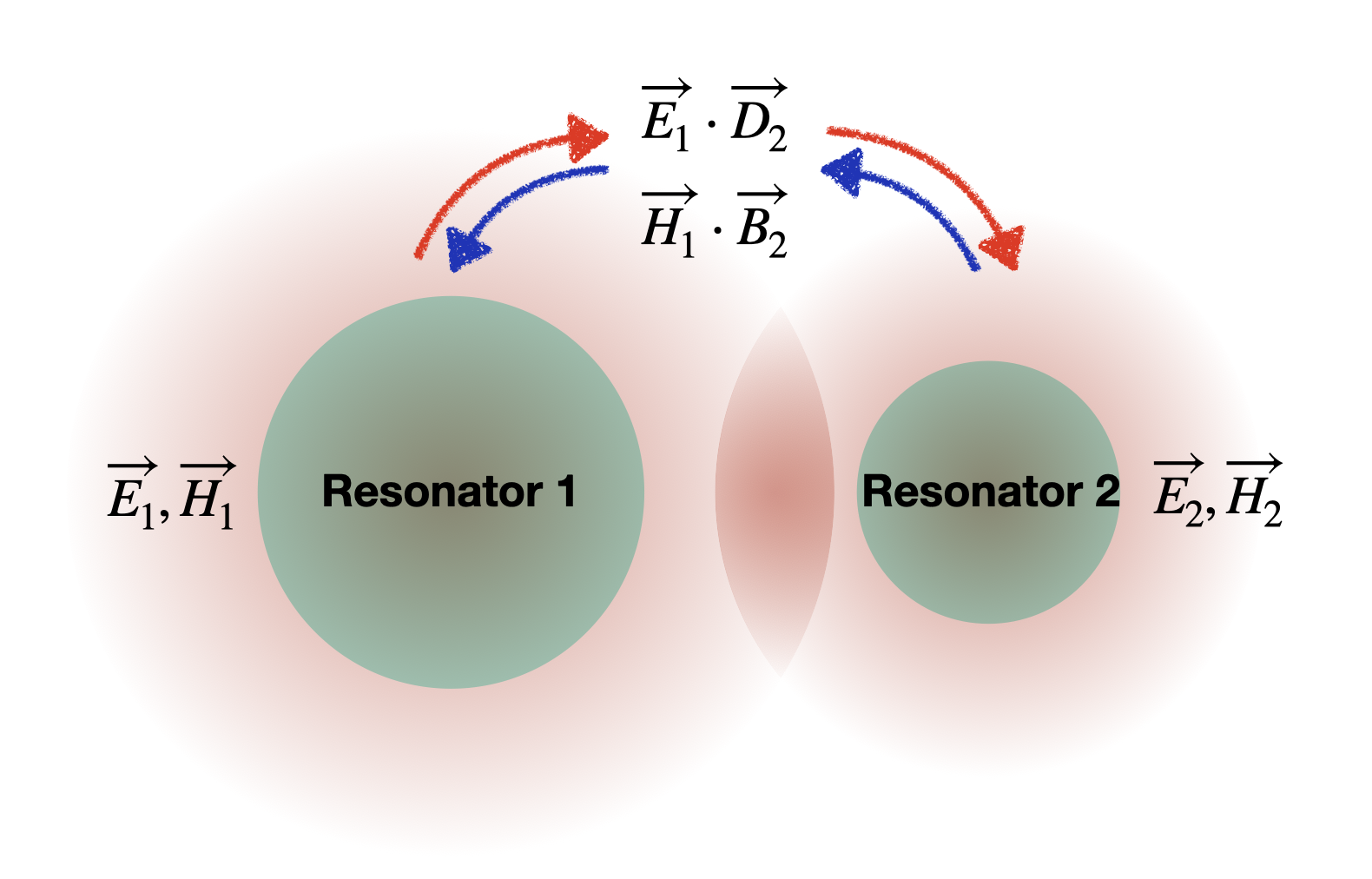}     
    \caption{Two coupled resonators exchanging energy via their overlapping fields.}
    \label{fig:CR}
\end{figure}

\begin{figure}[tbh!]
    \centering
    \subfloat[]{\includegraphics[clip,width=0.4\textwidth]{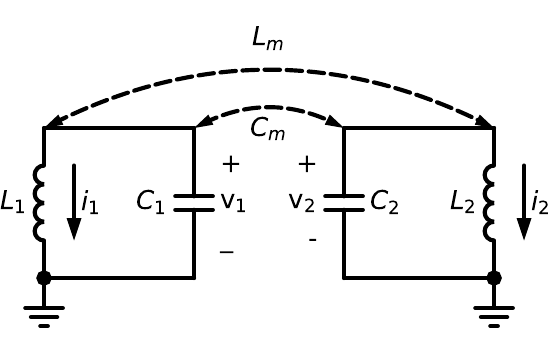}}

    \subfloat[]{\includegraphics[clip,width=0.45\textwidth]{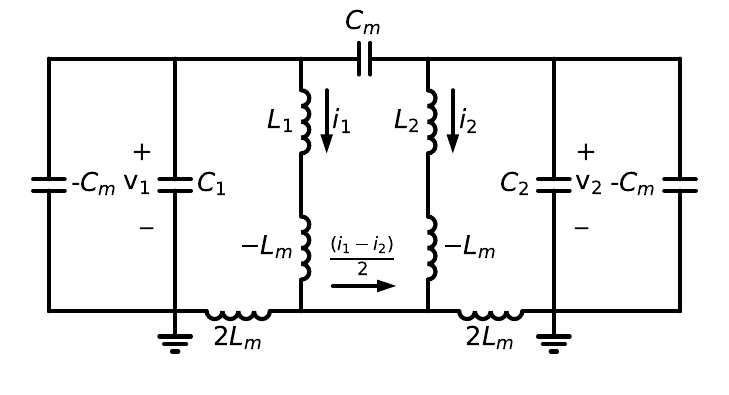}}
    \caption{Two coupled resonators. (a) and (b) are equivalent, described by (\ref{eq:9}) and \eqref{eq:10}.} \label{fig:coupled_resonators}
\end{figure}

Consider two coupled resonators as shown in Fig. \ref{fig:coupled_resonators}. 
Note that both circuits in Fig. \ref{fig:coupled_resonators} are equivalent, and are described by
\begin{gather}
    \mathbf{v} = \mathbf{L} \frac{d\mathbf{i}}{dt}\label{eq:9}, \\
    -\mathbf{i} = \mathbf{C} \frac{d\mathbf{v}}{dt}\label{eq:10},
\end{gather}
where $\mathbf{v} = \begin{pmatrix} v_1 & v_2 \end{pmatrix}^{\mathrm{t}}$ and $\mathbf{i} = \begin{pmatrix} i_1 & i_2 \end{pmatrix}^{\mathrm{t}}$ and
\begin{gather}
    \mathbf{L}=\begin{pmatrix} L_{1} & L_{m}\\ L_{m} & L_{2} \end{pmatrix}, \quad \mathbf{C}=\begin{pmatrix} C_{1} & -C_{m}\\ -C_{m} & C_{2} \end{pmatrix}.
\end{gather}

The negative capacitors/inductors in Fig. \ref{fig:coupled_resonators} are added to simplify the formulations; one can easily combine them with the resonators elements. 

Using \eqref{eq:12}, \eqref{eq:9}, and \eqref{eq:10}, after some algebra,
\begin{gather}
    \dot{a}_{r\pm}=\pm j\omega_{0r}a_{r\pm}+\frac{\left(\zeta_{C}\mp\zeta_{L}\right)}{2}\dot{a}_{s+}+\frac{\left(\zeta_{C}\pm\zeta_{L}\right)}{2}\dot{a}_{s-}, \label{eq:coupled_resonators}\\
     \zeta_L = \frac{L_m}{\sqrt{L_1 L_2}}, \quad \zeta_C = \frac{C_m}{\sqrt{C_1 C_2}}, \quad \omega_{0r}=\frac{1}{\sqrt{L_rC_r}},
\end{gather}
where $r,s \in \{1,2\}$ and $r\neq s$. Recasting (\ref{eq:coupled_resonators}) to isolate derivatives, 

\begin{equation}
    \begin{pmatrix}
    \dot{\mathbf{a}}_{1}\\
    \dot{\mathbf{a}}_{2}
    \end{pmatrix} = \begin{pmatrix}
    \mathbf{U}_0 & \mathbf{U}_g
    \\
    \mathbf{U}_g & \mathbf{U}_0 \end{pmatrix} \begin{pmatrix}
    j\omega_{01}\mathbf{a}_{1}\\
    j\omega_{02}\mathbf{a}_{2}
    \end{pmatrix} \label{eq:matrix_coupled_res}
\end{equation}
where
\begin{equation}
    \mathbf{a}_r = \begin{pmatrix} a_{r+} \\ a_{r-} \end{pmatrix}, \quad \mathbf{U}_0 = \begin{pmatrix} k_1 & -k_2 \\ k_2 & -k_1\end{pmatrix}, \quad
    \mathbf{U}_g = \begin{pmatrix} k_3 & -k_4 \\ k_4 & -k_3\end{pmatrix} \label{eq:coupled_res_matrix_defs}
\end{equation}
and
\begin{gather}
    k_{1}=\frac{2-\left(\zeta_{C}^{2}+\zeta_{L}^{2}\right)}{2(1-\zeta_{C}^{2})(1-\zeta_{L}^{2})}, \quad   k_{2}=\frac{\zeta_{C}^{2}-\zeta_{L}^{2}}{2(1-\zeta_{C}^{2})(1-\zeta_{L}^{2})}, \nonumber\\
    k_{3}=\frac{\left(1+\zeta_{C}\zeta_{L}\right)\left(\zeta_{C}-\zeta_{L}\right)}{2(1-\zeta_{C}^{2})(1-\zeta_{L}^{2})}, \quad     k_{4}=\frac{\left(1-\zeta_{C}\zeta_{L}\right)\left(\zeta_{C}+\zeta_{L}\right)}{2(1-\zeta_{C}^{2})(1-\zeta_{L}^{2})}. \label{eq:coupled_res_constants}
\end{gather}

It is common to use the following two approximations: 

 (a) ignore the second-order terms $\zeta_C^2$ and $\zeta_L^2$ since they both are  $\ll1.$ Therefore, 
 
\begin{equation}
    \mathbf{U}_0 \approx \begin{pmatrix} 1 & 0 \\ 0 & -1 \end{pmatrix}, \quad
    \mathbf{U}_g \approx \frac{1}{2}\begin{pmatrix} \zeta_C - \zeta_L & -(\zeta_C + \zeta_L) \\ \zeta_C + \zeta_L & \zeta_L - \zeta_C \end{pmatrix}. \label{eq:coupled_res_approximate}
\end{equation}

(b) ignoring the terms that couple $+$ and $-$ amplitudes, which is also known as RWA. This is justified if solutions to $a_{r+}$ and $a_{r-}$ have the form $\widetilde{a}_{r+}e^{j\omega_{0r}t}$ and
$\widetilde{a}_{r-}e^{-j\omega_{0r}t}$, respectively, where $\widetilde{a}_{r\pm}$
have  slow time variations compared to the exponential terms. 

Therefore, (\ref{eq:matrix_coupled_res}) reduces to
\begin{equation}
    \dot{a}_{r\pm} \approx \pm j \omega_{0r} a_{r\pm} \pm j \omega_{0s} \frac{\zeta_{C}-\zeta_{L}}{2} a_{s\pm}, \label{eq:coupled_res_approximate}
\end{equation}
Note that if $\zeta_C=-\zeta_L$, this is not an approximation anymore \cite{sank2024balanced}.
If $\zeta_C=\zeta_L$, there will be a small coupling between $a_{r\pm}$ and $a_{s\mp}$, which is worth exploring and is beyond the scope of this note.
The total energy in the system is
\begin{equation}
    W_{\mathrm{tot}}=\frac{1}{2}\left(\mathbf{v}^{\mathrm{t}}\mathbf{Cv}+\mathbf{i}^{\mathrm{t}}\mathbf{Li}\right).
\end{equation}
After some algebra, 
\begin{align}
    W_{\text{tot}} = & a_{1+}a_{1-} + a_{2+}a_{2-} -\frac{\zeta_C}{2}\left(a_{1+} + a_{1-}\right)\left(a_{2+} + a_{2-}\right) \nonumber \\
    & -\frac{\zeta_L}{2}\left(a_{1+} - a_{1-}\right)\left(a_{2+} - a_{2-}\right). \label{eq:W_tot}
\end{align}
Equation \eqref{eq:matrix_coupled_res} naturally satisfies energy conservation $\dot{W}_{\text{tot}} = 0$. Note that the energy is conserved only if we include the coupling terms in \eqref{eq:W_tot}. 

Another popular notation for the coupled resonators is to remove the $\omega_0$ coefficient in \eqref{eq:12} which leads to a symmetrical coupling term in \eqref{eq:coupled_res_approximate}. This is clarified in Appendix \ref{sec:coupled_res_symmetric}.

The coupling coefficients in (\ref{eq:coupled_res_approximate})  are the circuit equivalents of the right hand side of (\ref{eq:cc}), 
\begin{equation}
   \zeta_C=\frac{\int dv \, \varepsilon\Vec{E_{1}} \cdot \Vec{E_2}}{\sqrt{\int dv \, \varepsilon\abs{E_1}^2  \times\int dv \, \varepsilon \abs{E_2}^2}},
\end{equation}
\begin{equation}
   \zeta_L=\mp \frac{\int dv \, \mu\Vec{H_1} \cdot \Vec{H_2}}{\sqrt{\int dv \, \mu\abs{H_1}^2 \times \int dv \,\mu \abs{H_2}^2}}, \label{eq:magnetic_coupling}
\end{equation}
in which minus(plus) sign is for positive(negative) $L_m$ in \eqref{eq:9}.

Diagonalizing \eqref{eq:coupled_res_approximate} (i.e., looking for solutions as $a_{k\pm}=c_{k\pm}e^{j\omega t}$ where $c_{k\pm}$ is a constant) leads to the eigen frequencies  
\begin{equation}
    \omega_{1,2}=\frac{\left(\omega_{01}+\omega_{02}\right)}{2}\pm\frac{\sqrt{\left(\omega_{01}-\omega_{02}\right)^{2}+\left(\zeta_C-\zeta_L\right)^{2}\omega_{01}\omega_{02}}}{2}\label{eq:eigen}
\end{equation}
Re-organizing (\ref{eq:eigen}) \cite{bahl2001rf}, 
\begin{equation}
    \zeta_C-\zeta_L=\pm\left(\frac{\omega_{02}}{\omega_{01}}+\frac{\omega_{01}}{\omega_{02}}\right)\sqrt{\left(\frac{\omega_{2}^{2}-\omega_{1}^{2}}{\omega_{2}^{2}+\omega_{1}^{2}}\right)^{2}-\left(\frac{\omega_{02}^{2}-\omega_{01}^{2}}{\omega_{02}^{2}+\omega_{01}^{2}}\right)^{2}}\label{eq:kappa}
\end{equation}  
where  $\omega_{01,02}$  are the bare resonance frequencies and $\omega_{1,2}$ are the normal resonance frequencies of the coupled system.  This is the relation that is commonly used to extract the coupling coefficient, instead of (\ref{eq:cc}). 

By setting  $\omega_{01}=\omega_{02}$, (\ref{eq:kappa}) reduces to $\zeta_C-\zeta_L=\frac{\omega_2^2-\omega_1^2}{\omega_2^2+\omega_1^2}$, used in symmetric resonators. If the coupling is weak, the approximate relation $\zeta_C-\zeta_L=\frac{\omega_2-\omega_1}{\omega_{01}}$ can also be used.

\begin{figure}[b]
    \centering
    \includegraphics[width=0.5\textwidth]{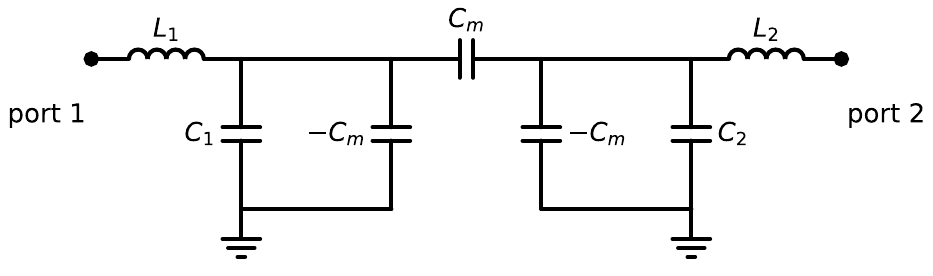}     
    \caption{Capacitively coupled resonators driven by two microwave  ports.}
    \label{fig:CR2}
\end{figure}

\begin{figure}[]
    \centering
    \includegraphics[width=0.5\textwidth]{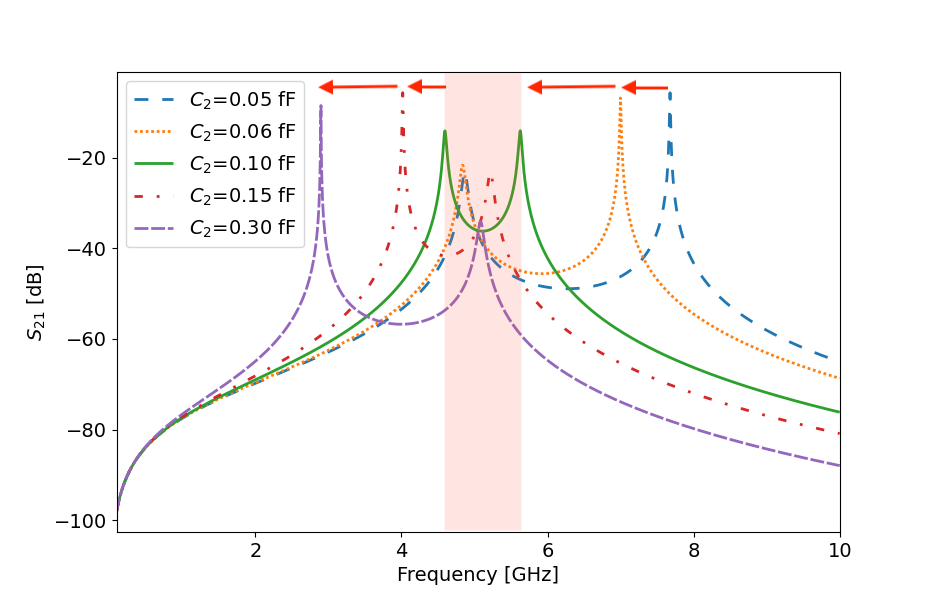}     
    \caption{Transmission through the coupled resonators of Fig. \ref{fig:CR2}. The peaks indicate the eigenfrequencies.  The sharper peak is associated with the $L_2C_2$ resonator, and the red arrows show its displacement as $C_2$ is varied. The two resonance modes are completely hybridized and indistinguishable when $L_1C_1=L_2C_2$. The avoided crossing region is shaded. $L_1=0.1\, \mu H$, $C_1=10\, fF$, $L_2=10\, \mu H$, and the port impedances are $50\, \Omega$.}
    \label{fig:AC}
\end{figure}

It is  evident from (\ref{eq:eigen}) that the normal modes of the coupled resonators become farther apart in frequency as the coupling coefficient increases.
As an illustration, consider two capacitively coupled resonators as shown in Fig. \ref{fig:CR2}, in which two microwave ports with low impedance are used to connect the inductors  to the ground. This allows us to examine the normal modes of the system using its transmission response (i.e. $S_{21}$), shown in Fig. \ref{fig:AC}.
The inductors in Fig. \ref{fig:CR2} have different values so that the bare modes of the two resonators are discernible in the transmission spectrum.
The resonator with the higher inductance $(L_2)$ has the sharper peak in Fig. \ref{fig:AC}.
Decreasing this resonator's frequency (by increasing $C_2$) brings the normal modes closer together until they hybridize and have equal peaks (the green curve in Fig. \ref{fig:AC}).
This is where $\omega_{01}=\omega_{02}$ in (\ref{eq:eigen}), and $\omega_2-\omega_1=2g_1$.
Decreasing $C_2$ separates the normal modes further again.
This behavior is known as the avoided crossing and has numerous applications in sensing, microwave devices, antennas, etc.
A common equivalent statement is that any added coupling between degenerate modes would lift their degeneracy, i.e. any coupling hybridizes the modes and pushes their frequencies away from each other.

\begin{figure}[tbh]
    \centering
    \includegraphics[width=0.5\textwidth]{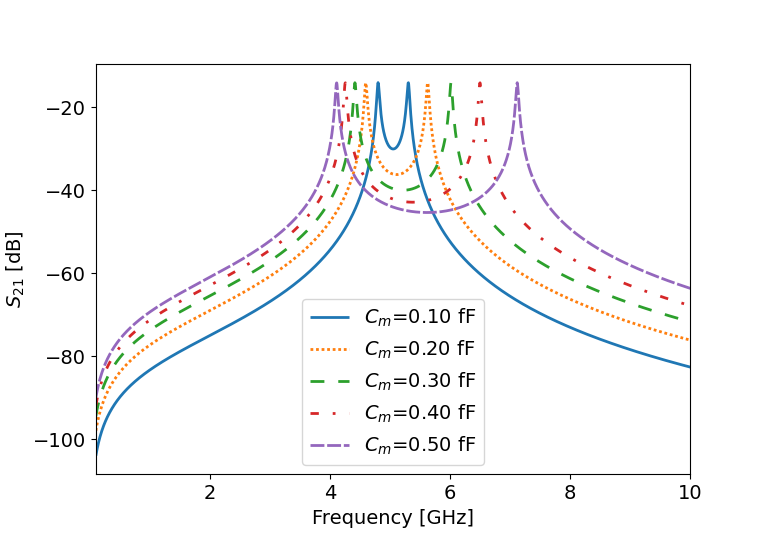}     
    \caption{The avoided crossing, the frequency distance between the peaks, as the coupling strength varies. }
    \label{fig:AC2}
\end{figure}

As mentioned earlier, the minimum frequency separation of the normal modes, a.k.a. the avoided region, is proportional to the coupling strength, $g_1$.
Figure \ref{fig:AC2} shows the transmission spectra of the hybridized modes when the coupling capacitor is increased. 
It is worth mentioning that in the presence of both gain and loss, an exceptional point of degeneracy can be created between two coupled resonators (modes can cross each other).
This has gained a lot of interest in sensing applications, recently.  

In time domain, if one of the coupled resonators is excited by a delta function $\delta(t)$, both normal modes will be excited.
As the system evolves in time, part of the system`s energy oscillates between the two resonators.
If the partial frequencies are equal, exciting one of the resonators by a delta function will excite both hybridized modes equally.
As they evolve in time the entire energy of the system oscillates between the two resonators.
The frequency of this oscillation is determined by the coupling strength. See \cite{krage1970characteristics,esfahani2007new} for more information about resonator-resonator couplings.

\section{Uniformly coupled transmission lines}
Coupled transmission lines analysis has applications in designing qubits' readout lines, as well as minimizing the unwanted couplings in the device. In the followings, the eigen mode analysis is reviewed, which is useful in designing couplers between the readout resonators and the feedline (e.g. in a multiplexed readout system).
The theory of weakly coupled transmission lines is also briefly reviewed in Appendix \ref{sec:Weakly_coupled_lines}. It has applications in calculating the unwanted couplings between parallel lines.
The discussion here is limited to uniform symmetrical transmission lines. Both above theories are vastly developed in microwave engineering, beyond uniform lines \cite{Malherbe1988}.

\leveldown{Eigen mode analysis}
\begin{figure}
    \centering
    \includegraphics[width=0.4 \textwidth]{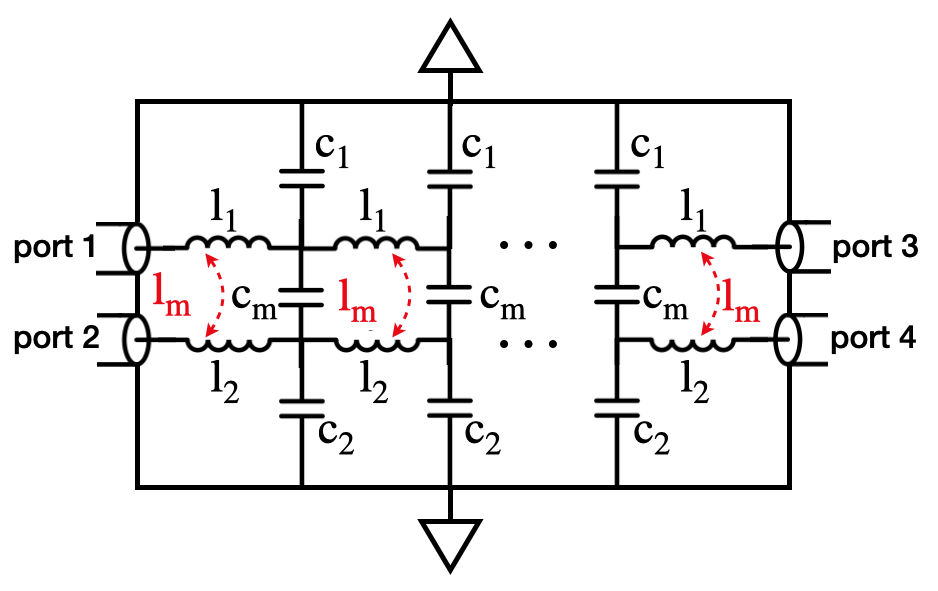}
    \caption{Parallel transmission lines represented by a 4-port network.}
    \label{fig:PTL}
\end{figure}
Consider a pair of coupled transmission lines, as shown in Fig. \ref{fig:PTL}. Then, \cite{ElecWaves&AntennasOrfanidis}
\begin{gather}
    \frac{\partial \mathbf{v}}{\partial z} = - \mathbf{L}_d \frac{\partial \mathbf{i}}{\partial t},\label{eq:telegrapher-1} \\ 
    \frac{\partial \mathbf{i}}{\partial z} = -  \mathbf{C}_d \frac{\partial \mathbf{v}}{\partial t} \label{eq:telegrapher-2}
\end{gather}
where $\mathbf{v} = \begin{pmatrix} v_1 & v_2 \end{pmatrix}^{\mathrm{t}}$ and $\mathbf{i} = \begin{pmatrix} i_1 & i_2 \end{pmatrix}^{\mathrm{t}}$ are voltages and currents of the transmission lines, respectively, and 
\begin{equation}
     \mathbf{L}_d =\begin{pmatrix} l_1 & l_m \\ l_m & l_2 \end{pmatrix},\quad  \mathbf{C}_d =\begin{pmatrix} c_1 & -c_m \\ -c_m & c_2 \end{pmatrix},
\end{equation} 
are the inductance and capacitance density matrices, respectively.

For simplicity, let us consider the symmetric case, which is often designed for, $l_1 = l_2 = Z_0/v_{ph}$ and $c_1 = c_2 = 1/(v_{ph} Z_0)$, where $v_{ph}$ is the phase velocity of each line in isolation and $Z_0$ the characteristic impedance.  In this case $\mathbf{L}_d$ and $\mathbf{C}_d$ are both of the form
\begin{equation}
    k_{\mathbf{I}} \mathbf{I} + k_{\mathbf{X}} \mathbf{X}, \quad \mathbf{X} = \begin{pmatrix} 0 & 1 \\ 1 & 0 \end{pmatrix} \label{eq:symmetric}
\end{equation}
where $k_i$ are constants. All matrices of this form, and all functions of matrices of this form, commute with each other, greatly simplifying our algebraic efforts. For even further ease, we can transform all matrices of the from \eqref{eq:symmetric} to diagonal form with use of the Hadamard gate 
\begin{equation}
    \mathbf{H} = \frac{1}{\sqrt{2}} \begin{pmatrix} 1 & 1 \\ 1 & -1 \end{pmatrix},
\end{equation}
which naturally separates the system into even and odd modes.

The amplitudes
\begin{align}
    \boldsymbol{\alpha}_\pm = \frac{\mathbf{I} - \mathbf{\Gamma}}{2}\left(\mathbf{v} \pm \mathbf{Z i}\right), \quad \mathbf{\Gamma} = (\mathbf{Z} - Z_0 \mathbf{I})(\mathbf{Z} + Z_0 \mathbf{I})^{-1},
\end{align}
where $\mathbf{Z}=\sqrt{\mathbf{L}_d\mathbf{C}_d^{-1}}$, block diagonalize \eqref{eq:telegrapher-1}-\eqref{eq:telegrapher-2},
leading to
\begin{align}
    \frac{\partial\boldsymbol{\alpha}_\pm}{\partial z} = \mp \sqrt{\mathbf{L}_d\mathbf{C}_d} \frac{\boldsymbol{\alpha}_\pm}{\partial t}.
\end{align}
The reflection matrix $\mathbf{\Gamma}$ is included for later algebraic convenience. For further ease we will work with in the Fourier basis $e^{j\omega t}$. Then,
\begin{align}
    \frac{d\boldsymbol{\alpha}_\pm}{d z} = \mp j \mathbf{B} \boldsymbol{\alpha}_\pm ,
\end{align}
where
\begin{gather}
    \mathbf{HBH} = \begin{pmatrix} \beta_+ & 0 \\ 0 & \beta_- \end{pmatrix}, \quad \mathbf{HZH} = \begin{pmatrix} Z_+ & 0 \\ 0 & Z_- \end{pmatrix}, \nonumber\\
    \beta_\pm = \beta \sqrt{(1 \mp \zeta_C)(1 \pm \zeta_L)}, \quad Z_\pm = Z_0\sqrt{\frac{1 \pm \zeta_L}{1 \mp \zeta_C}}, \nonumber \\
    \zeta_L = \frac{l_m}{\sqrt{l_1 l_2}}, \quad \zeta_C = \frac{c_m}{\sqrt{c_1 c_2}},
\end{gather}
with $\beta = \omega/v_{ph}$. 

Suppose then that we wish to find the $\mathbf{S}$-matrix for incoming and outgoing power waves. We define
\begin{align}
    \mathbf{v} = \mathbf{v}_+ + \mathbf{v}_-, \quad Z_0 \mathbf{i} = \mathbf{v}_+ - \mathbf{v}_-.
\end{align}
Expressing in terms of the aforementioned reflection matrix $\mathbf{\Gamma}$,
\begin{equation}
    \begin{pmatrix} \boldsymbol{\alpha}_+ \\ \boldsymbol{\alpha}_- \end{pmatrix} = \begin{pmatrix} \mathbf{I} & -\mathbf{\Gamma} \\ -\mathbf{\Gamma} & \mathbf{I} \end{pmatrix} \begin{pmatrix} \mathbf{v}_+ \\ \mathbf{v}_- \end{pmatrix}.
\end{equation}
If we take a coupler length $\ell$, then we can relate
\begin{align}
    \begin{pmatrix} \mathbf{v}_-(\ell) \\ \mathbf{v}_+(\ell) \end{pmatrix} = & \left[\mathbf{I} \otimes (\mathbf{I} - \mathbf{\Gamma}^2)^{-1}\right] \begin{pmatrix} \mathbf{I} & \mathbf{\Gamma} \\ \mathbf{\Gamma} & \mathbf{I} \end{pmatrix} \nonumber \\
    & \times \begin{pmatrix} e^{-j \mathbf{B} \ell} & \mathbf{0} \\ \mathbf{0} & e^{j \mathbf{B} \ell} \end{pmatrix} \begin{pmatrix} \mathbf{I} & -\mathbf{\Gamma} \\ -\mathbf{\Gamma} & \mathbf{I} \end{pmatrix} \begin{pmatrix} \mathbf{v}_+(0) \\ \mathbf{v}_-(0) \end{pmatrix}.
\end{align}
Now noting \eqref{eq:power-to-tl} and rearranging, the power wave $\mathbf{S}$-matrix for the four-port network is given by
\begin{equation}
    \mathbf{S} = \left[\mathbf{I} \otimes \left(e^{j\mathbf{B}\ell} - \mathbf{\Gamma}^2 e^{-j\mathbf{B}\ell}\right)^{-1}\right]\begin{pmatrix} j2\mathbf{\Gamma} \sin \mathbf{B}\ell & \mathbf{I} - \mathbf{\Gamma}^2 \\ \mathbf{I} - \mathbf{\Gamma}^2 & j2\mathbf{\Gamma} \sin \mathbf{B}\ell\end{pmatrix}.
\end{equation}
Let us consider a coupler design which features no reflection; that is, $\text{diag}(\mathbf{S}) = 0$. Equivalently, this means $\text{diag}(\mathbf{\Gamma} \sin \mathbf{B}\ell) = 0$, which yields the condition
\begin{equation}
    \frac{Z_+ - Z_0}{Z_+ + Z_0} \sin\beta_+ \ell + \frac{Z_- - Z_0}{Z_- + Z_0}\sin\beta_- \ell = 0.
\end{equation}
There are two cases which satisfy this regardless of $\ell$. The first is the ``forward-coupler" where $Z_+ = Z_- = Z_0$, which is impedance-matched hence $\mathbf{\Gamma} = \mathbf{0}$. In terms of couplings, this is when $\zeta_L = -\zeta_C$. This case reduces simply to
\begin{equation}
    \mathbf{S} = \begin{pmatrix} \mathbf{0} & e^{-j\mathbf{B}\ell} \\  e^{-j\mathbf{B}\ell} & \mathbf{0} \end{pmatrix}.
\end{equation}
Specifically looking at the power transfer between lines,
\begin{equation}
   |\mathbf{S}_{41}| = \left|\sin\left(\frac{\beta_+ - \beta_-}{2}\ell\right)\right| = |\sin\left(\zeta_C \beta \ell\right)|.
\end{equation}

The second case is the ``backward-coupler" with the conditions $\beta_+ = \beta_-$ and $Z_+ Z_- = Z_0^2$. In terms of couplings, this is when $\zeta_L = \zeta_C$. Then
\begin{equation}
    \mathbf{S} = \frac{1}{\sqrt{1-\zeta^2}\cos \theta + j \sin \theta}\begin{pmatrix} 
    (j\zeta\sin\theta)\mathbf{X} & \sqrt{1-\zeta^2}\mathbf{I} \\ 
    \sqrt{1-\zeta^2}\mathbf{I} & (j\zeta\sin\theta)\mathbf{X} \end{pmatrix},
\end{equation}
where $\theta = \beta\ell\sqrt{1-\zeta^2}$ and we define the dimensionless coupling
\begin{equation}
    \zeta = \frac{Z_+ - Z_-}{Z_+ + Z_-} = \zeta_L = \zeta_C.\label{eq:zeta}
\end{equation} 
This is also known as the voltage coupling coefficient. 
The power transfer between lines is then characterized by
\begin{equation}
   |\mathbf{S}_{21}| = \frac{\zeta |\sin\theta|}{\sqrt{1 - \zeta^2 \cos^2 \theta}}. \label{eq:S21}
\end{equation}

The same formulations can be obtained by considering  scattering matrices of the transmission lines, as summarized in Appendix  \ref{sec:EMA_coupled_lines}.
For more detailed information about asymmetrical directional couplers see \cite{sellberg1990formulas}. 

As an example, consider two parallel identical coplanar waveguide lines, without the ground in between. This is clarifed in the insert of Fig. \ref{fig:coupler_param}. The metals are assumed to be perfect conductors with the thickness of 400 nm, and the gaps are all fixed at $2\, \mu$m. Silicon is used as the substrate with the permittivity of 11.9.   
% \begin{figure}
%     \centering
%     \includegraphics[width=0.3 \textwidth]{Parallel_TLs.png}
%     \caption{Coupled CPW lines with no ground in between.}
%     \label{fig:Parallel_TL}
   
% \end{figure}

The trace width W is varied to minimize  $\left|\beta_+-\beta_-\right|$. As Fig. \ref{fig:coupler_param} shows, $\left|\beta_+-\beta_-\right|/\left|\beta_++\beta_-\right|$ is less than 2\% in the considered W range, which indicates this geometry inherently leads to a balanced coupler. This is because the effective permittivities of the even and odd modes are almost equal, if the metal thickness is small enough. 

\begin{figure}
    \centering
    \includegraphics[width=0.4 \textwidth]{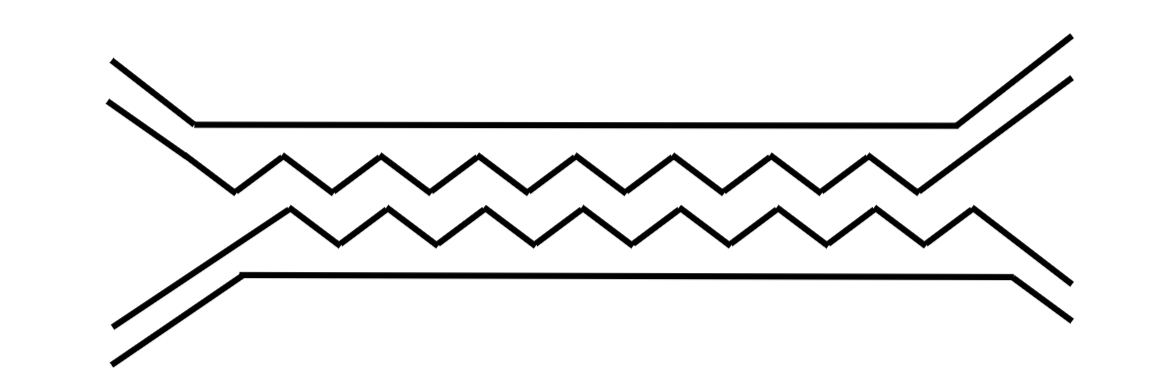}
    \caption{Increasing the fringe capacitance between the lines.}
    \label{fig:Wiggly_lines}
\end{figure}

However, in order to have a directional coupler, $Z_+Z_-=Z_0^2$ also needs to be satisfied. 
Fig. \ref{fig:coupler_param} shows that the trace width of $3\, \mu$m satisfies this condition. Since the metal thickness is not zero, there is a trade off between the impedance matching and the coupling balance in order to achieve a directive coupler. 
The voltage coupling coefficient \eqref{eq:zeta} is also shown in Fig. \ref{fig:coupler_param}. It  increases with W, as expected. Some possible methods to improve this coupler's directionality are changing the metal thickness, changing the dielectric between the traces, or increasing the fringe capacitance between traces by using ``wiggly lines'' as shown in Fig. \ref{fig:Wiggly_lines}.
\newlength\imagewidth
\newlength\imagescale
\begin{figure}
    \centering
    \pgfmathsetlength{\imagewidth}{\linewidth}%
    \pgfmathsetlength{\imagescale}{\imagewidth/524}%
    \begin{tikzpicture}[x=\imagescale,y=-\imagescale]
        \node[anchor=north west] at (0,0) {\includegraphics[width=\imagewidth]{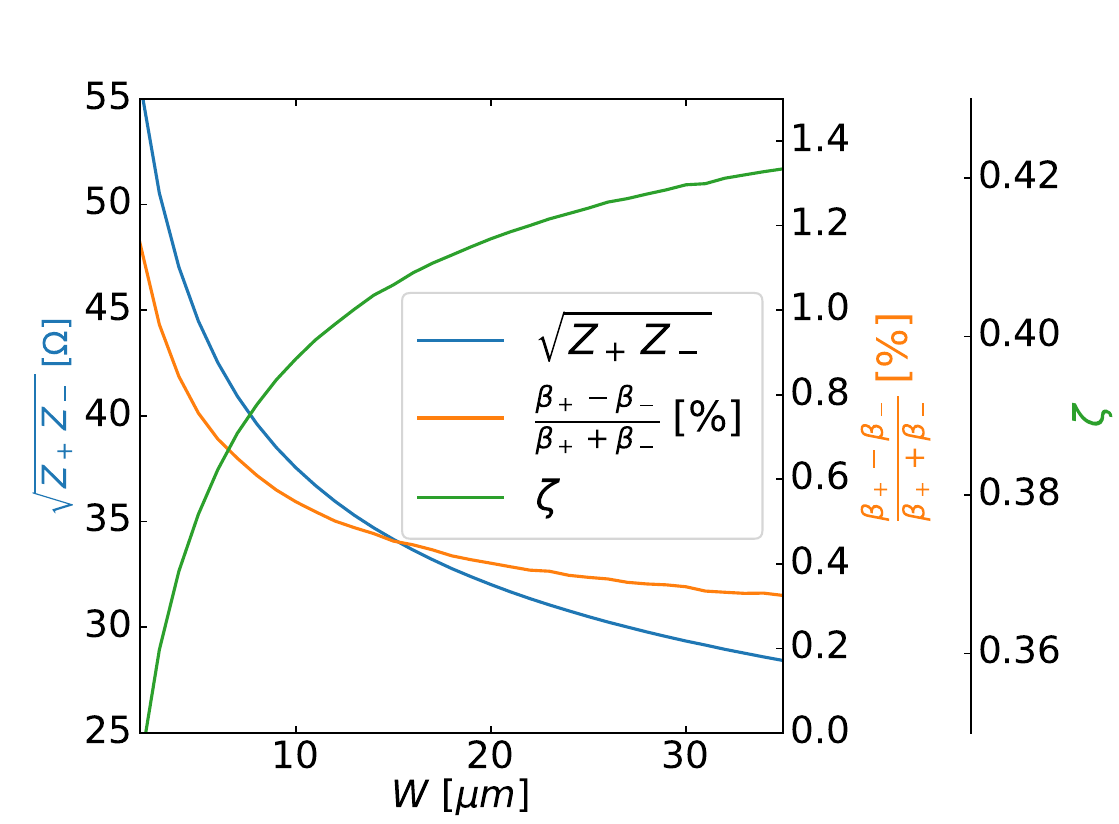}};
        \node[anchor=north west] at (90,47) {\includegraphics[width=0.2\imagewidth]{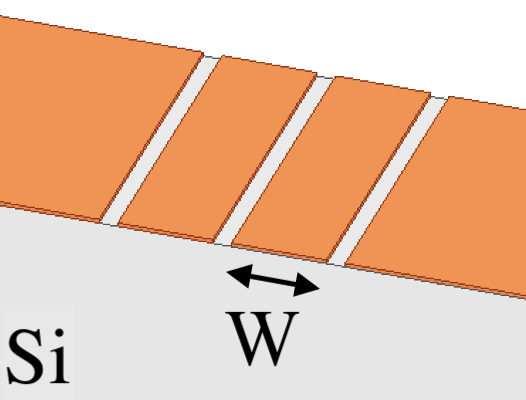}};
    \end{tikzpicture}
    \caption{Parameters of symmetrical coupled coplanar waveguide lines with no ground in between, as a function of the trace width.}
     \label{fig:coupler_param}
\end{figure}
% \begin{figure}
%     \centering
%     \includegraphics[width=0.5 \textwidth]{coupler_all.pdf}
%     \caption{ The coupler`s impedance, voltage coupling coefficient, and imbalance as a function of W.}
%     \label{fig:coupler_param}
% \end{figure}

\section{Couplers in distributed resonators}

\begin{figure}[h]
    \centering
    \subfloat[]{\includegraphics[clip,width=0.4\textwidth]{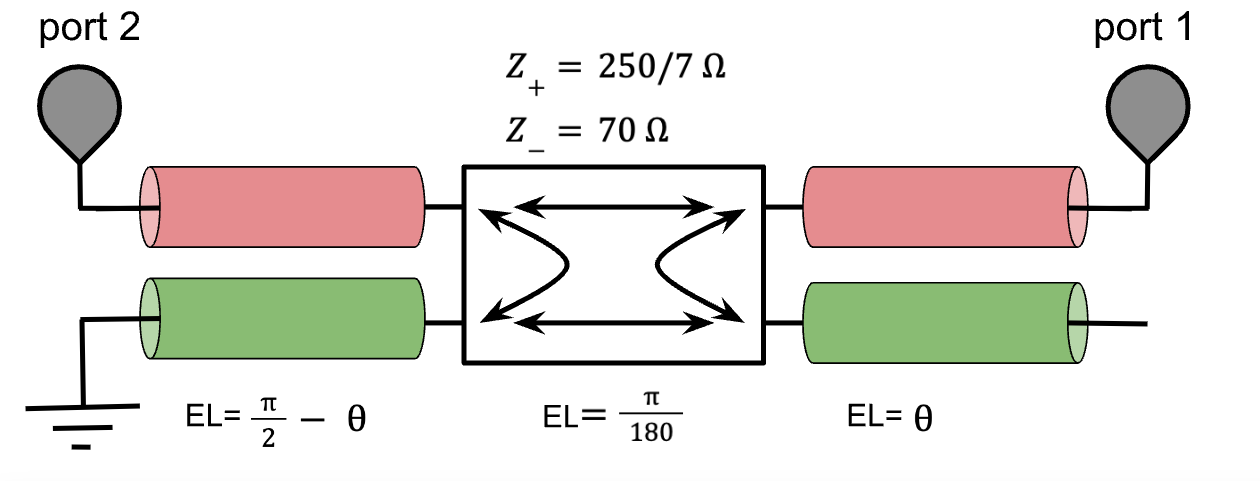}}
    
    \subfloat[]{\includegraphics[clip,width=0.4\textwidth]{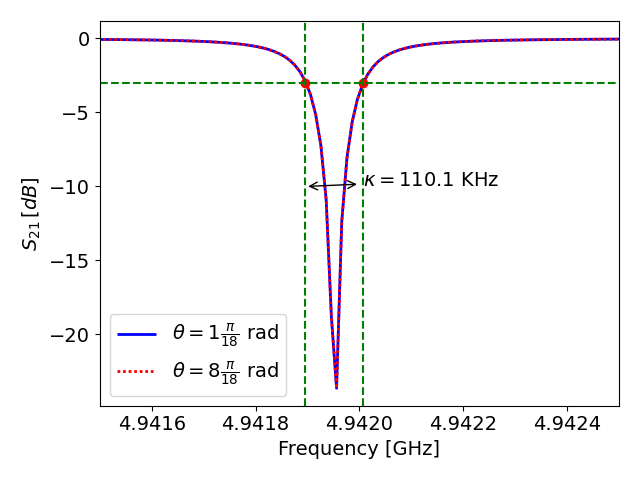}} 
    \caption{(a) a $\lambda/4$ resonator coupled to a transmission line. "EL" is the electrical length at $5 \, \mathrm{GHz}$. The characteristic impedance of all ports and transmission lines is $50\,\Omega$, (b) the scattering response of the system.}
    \label{fig:coupler_tranmissive}
\end{figure}

Parallel transmission line couplers are very common in coupling distributed resonators to their feeding transmission lines in superconducting devices. Consider a $\lambda/4$ resonator coupled to a transmission line as shown in Fig. \ref{fig:coupler_tranmissive}. The coupler's even and odd impedances are set to realize a backward-directional coupler with the electrical length of 1 degree at the frequency of $5\,\mathrm{GHz.}$

\begin{figure}[h]
    \centering
    \includegraphics[width=0.5\textwidth]{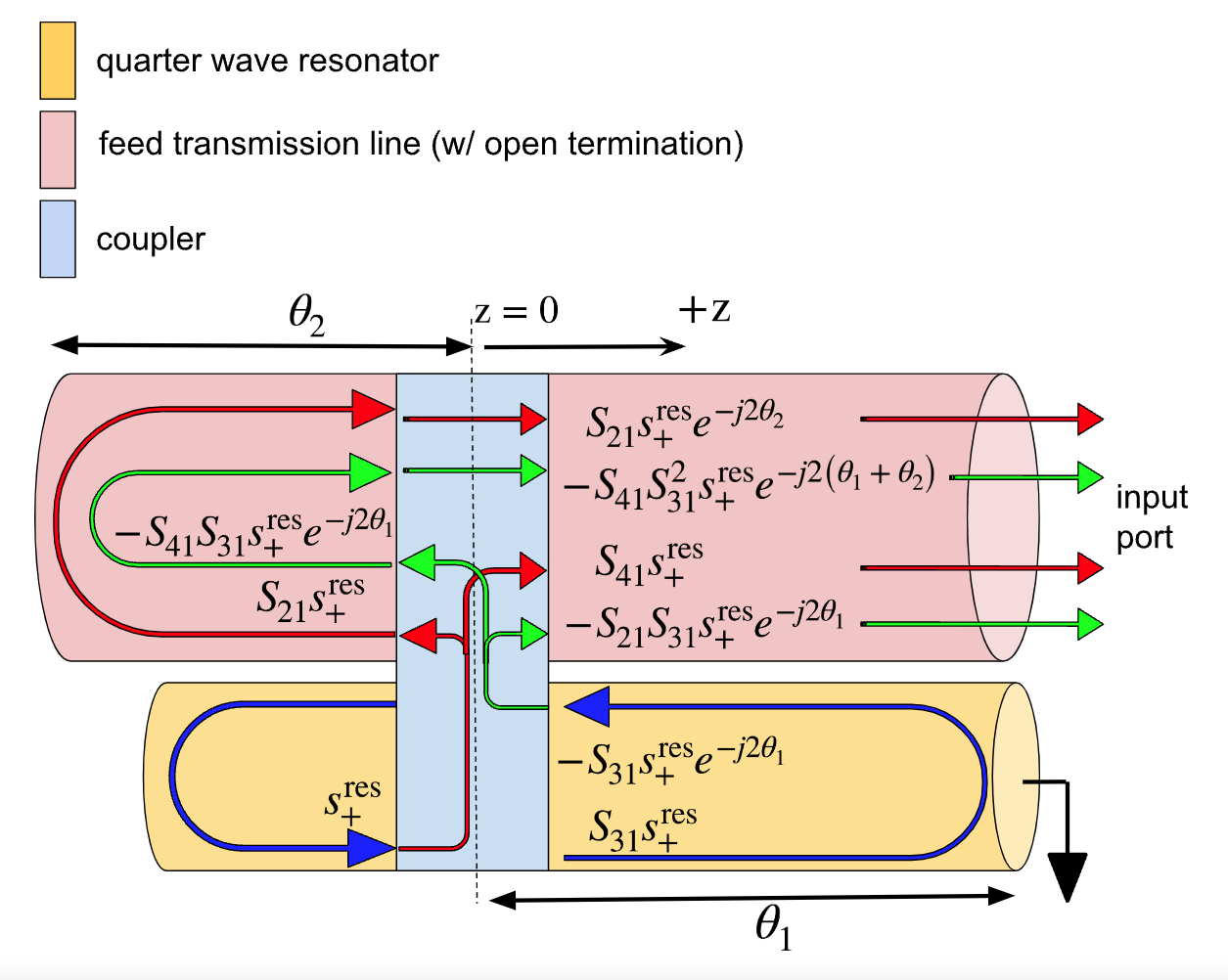}    
    \caption{An open ended transmission line coupled to a $\lambda/4$ resonator. The specified power waves are at $z=0$.}
    \label{fig:phase_matching}
\end{figure}

Fig. \ref{fig:coupler_tranmissive} also shows the transmission signal as the coupler is moved from near the open end ($\theta=10^o$) to near the short end ($\theta=80^o$). The coupling coefficient remains unchanged ($\simeq 110$ kHz) in both cases, which is the result of using backward-directional coupler. In order to relate $\kappa$ to the directional coupler's scattering parameters, assume the circulating power wave $s_+^{\mathrm{res}}$ in the resonator. It creates power waves $s_{2-}$ and $s_{3-}$ in the transmission line travelling towards ports 2 and 3, respectively. Therefore, the total energy decay rate of the resonator is 
\begin{equation}
    \frac{dW}{dt}=|s_{2-}|^2+|s_{3-}|^2=2|S_{21}|^2|s_+^{\mathrm{res}}|^2,\label{eq:transmissive_p_waves}
\end{equation}
where $S_{21}$ of the directional coupler is defined in \eqref{eq:S21}. Using, \eqref{eq:powerwave_quarterwave} and \eqref{eq:transmissive_p_waves},

\begin{equation}
   \kappa= \frac{dW/dt}{W}=4f_0|S_{21}|^2\quad \mathrm{\frac{rad}{s}}.\label{eq:transmissive_decay_rate}
\end{equation}
Note that the port definitions for for $S_{21}$ are based on Fig. \ref{fig:PTL}.
As a more general example, consider a $\lambda/4$ resonator coupled to a transmission line with open termination as shown in Fig. \ref{fig:phase_matching}. Suppose the resonator is energized with the circulating power wave $s^{\mathrm{res}}_+$. The phase reference  for $s^{\mathrm{res}}_+$ is at the coupler's $\mathrm{z}$ (i.e., the coupler is at $z=0$.)
Each cycle of $s^{\mathrm{res}}_+$ in the resonator generates four outgoing power waves in the transmission line, as clarified in Fig. \ref{fig:phase_matching}. Note that the power wave aquires a $\pi$ phase shift upon reflection from short. The definitions of the coupler`s ports are as Fig. \ref{fig:PTL}.
In most practical applications, the electrical length of the coupler is small, leading to $S_{31}\simeq1$ (see appendix \ref{sec:EMA_coupled_lines}). If the coupler is also very directional, $S_{41}=0$, and $\theta_2-\theta_1=\frac{\pi}{2}$,  
\begin{equation}
    \kappa= \frac{dW/dt}{W}=4|S_{21}|^2|s_+^{\mathrm{res}}|^2=8f_0|S_{21}|^2\quad \mathrm{\frac{rad}{s}}.\label{eq:kappa_reflective}
\end{equation}
This is twice the transmissive decay rate \eqref{eq:transmissive_decay_rate} and is also independent of the coupler's location.  Similarly, choosing $\theta_1=\theta_2$ leads to zero coupling between the transmission line and the resonator. 

\begin{figure}[h]
    \centering
    \subfloat[]{\includegraphics[clip,width=0.4\textwidth]{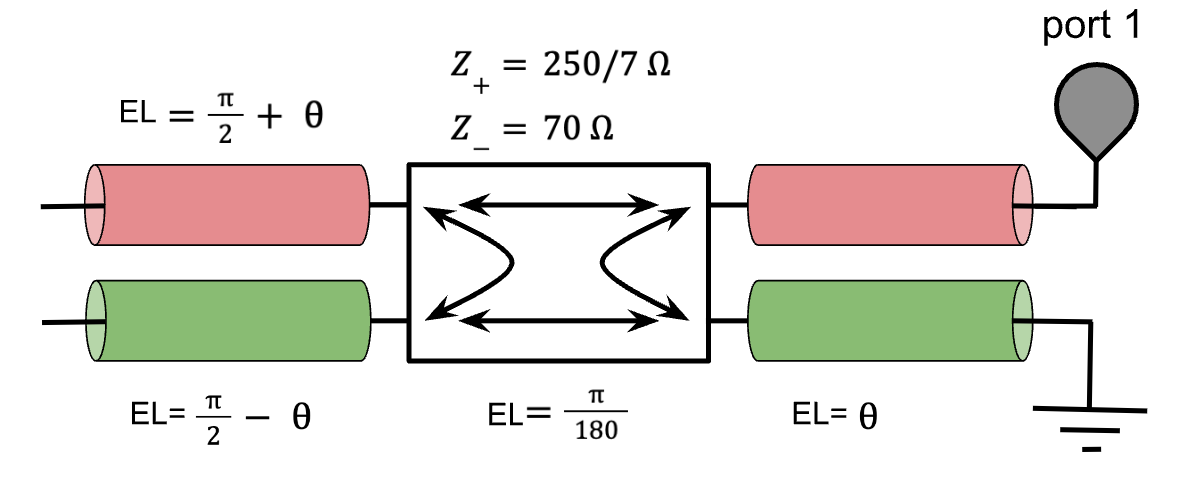}}
    
    \subfloat[]{\includegraphics[clip,width=0.4\textwidth]{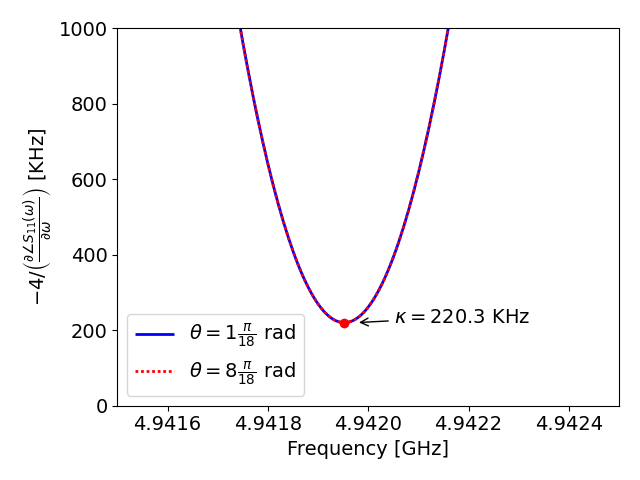}} 
    \caption{(a) A $\lambda/4$ resonator coupled to a single ended transmission line, (b) the reflection response of the system.}
    \label{fig:coupler_phase_matched}
\end{figure}

To verify \eqref{eq:kappa_reflective}, consider the circuit shown in Fig. \ref{fig:coupler_phase_matched} along with its Spice simulation result. As expected, the decay rate of the resonator is $\simeq 220$ kHz for different values of $\theta$, which is twice the transmissive example. 

Using \eqref{eq:S21} and \eqref{eq:kappa_reflective} for  parameters in Fig. \ref{fig:coupler_phase_matched},
\begin{equation}
    |S_{21}|=5.91\times 10^{-3};\quad \kappa =220.1 \,\mathrm{KHz}.\label{eq:calculated_kappa}
\end{equation}

\begin{figure}[h]
    \centering
    \subfloat[$Z_0^{even}$ and $Z_0^{odd}$ are varied while $\sqrt{Z_0^{even}Z_0^{odd}}=50$ is maintained. The coupler's electrical length is 1 degree at 5 GHz, and $\theta=45^o$.  ]{\includegraphics[clip,width=0.4\textwidth]{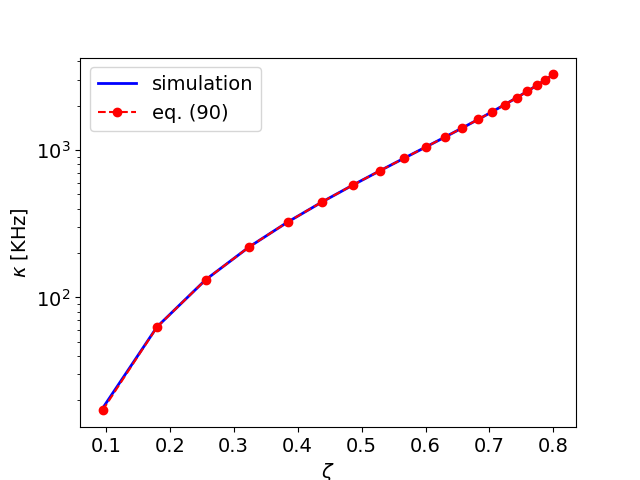}}
    
    \subfloat[The coupler's electrical length is  varied. $\theta=45^o$, $z_0^{even}=70\,\Omega$, and $z_0^{odd}=250/7\,\Omega$.]{\includegraphics[clip,width=0.4\textwidth]{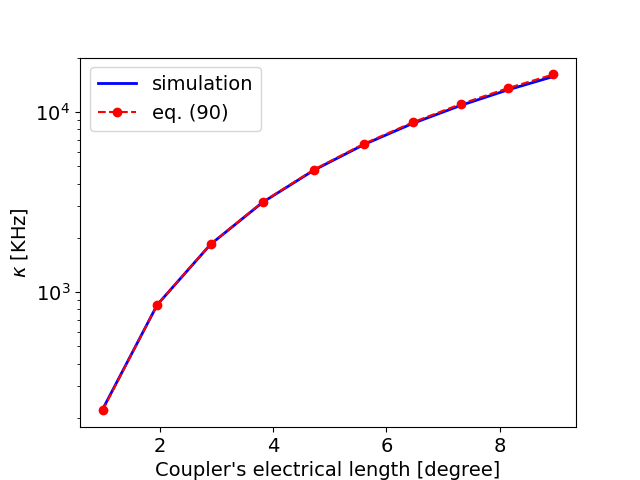}}
    \caption{$\kappa$ in Fig. \ref{fig:coupler_phase_matched} as the coupler's parameters are varied.}
    \label{fig:kappa_Comparisons}
\end{figure}

Comparing (\ref{eq:calculated_kappa}) with the $220.3\, \mathrm{kHz}$ from the circuit simulation, shown in Fig. \ref{fig:coupler_phase_matched}, the discrepancy is less than $ 0.1\% $. This means the approximations used in the analysis are sufficient for this range of frequencies and couplings. For instance, $a_-$ in the resonator was assumed to have no coupling to the transmission line's power waves. Comparisons between (\ref{eq:kappa_reflective}) and Spice simulations of the geometry in Fig. \ref{fig:coupler_phase_matched} are shown in Fig. \ref{fig:kappa_Comparisons}. As expected, the two approaches are in excellent agreement for different coupler parameters. 

\section{Conclusion}
The  electromagnetic  couplings  between resonators and transmission lines were discussed. The common approximations used in defining the system's equation of motion were clarified. Coupled transmission lines and their inclusion in distributed resonators were discussed.

\section*{Acknowledgment}
The Authors would like to thank Anthony Megrant, Daniel Sank, and Alexander Korotkov for the constructive discussions and comments. 

\appendices
\section{Relation between (\ref{eq:12}) and quantized fields}\label{sec:quantized fields in resonators}

In the  (second) quantization of the fields of a single mode in a  resonator, the coefficients  $\Vec{C}_{\mathrm{E}}(r)$ and $\Vec{C}_{\mathrm{H}}(r)$ are  properly chosen such that \cite{gerry2023introductory}
\begin{equation}
     \Vec{E}\left(t,r\right)=\Vec{C}_{\mathrm{E}}(r) q\left(t\right); \quad \Vec{H}\left(t,r\right)=\Vec{C}_{\mathrm{H}}(r) \dot{q}\left(t\right)=\Vec{C}_{\mathrm{H}}(r) p\left(t\right), \label{eq:second_quantization}
\end{equation}
and the classical field energy (Hamiltonian) of the mode is 
\begin{equation}
    H=\frac{1}{2}\left(p(t)^2+\omega_0^2q(t)^2\right).\label{eq:Hamiltonian_1D_res}
\end{equation}
This is equivalent to Hamiltonian of a harmonic oscillator of unit mass, indicating p and q are canonical variables. Elevating them to operators and imposing the canonical commutation relation $[\hat{q},\hat{p}]=i\hbar \hat{I}$ leads to the quantized fields and the definition of the annihilation operator as 
\begin{equation}
    \hat{a}=\frac{1}{\sqrt{2\hbar\omega_0}}\left(\omega_0 \hat{q}+j\hat{p}\right).\label{eq:annihilation_fields}
\end{equation}
As an example, consider a parallel plate transmission line along the z-axis with perfectly conducting walls at $z=0$ and $z=\ell$. It forms a 1D resonator with the electric and magnetic fields of its lowest frequency mode given by \cite{gerry2023introductory}
\begin{gather}
    E_{x}\left(z,t\right)=\omega_0\sqrt{\frac{2}{\varepsilon_{0} dw\ell}}q\left(t\right)\sin\left(\frac{\pi z}{L}\right), \label{eq:E_x} \\
   H_{y}\left(z,t\right)=\sqrt{\frac{2}{\mu_{0}dw\ell}}p\left(t\right)\cos\left(\frac{\pi z}{L}\right),\label{eq:H_y}
\end{gather}
where $d$ is the plates distance and $w$ is the effective width of the plates and $w/d$ is sufficiently large so that fringing fields can be ignored. The frequency of the resonator also satisfies  $\omega_0\sqrt{\mu_{0}\varepsilon_{0}}=\pi/\ell$. The coefficients in (\ref{eq:E_x}) and (\ref{eq:H_y}) are chosen such that (\ref{eq:Hamiltonian_1D_res}) is satisfied. 

Let us define the voltage and current in the equivalent LC circuit of the resonator as 
\begin{gather}
   v\left(t\right)=\left(\frac{\varepsilon_0 w d}{C_{\text{eff}}}\int_{0}^{\ell}dz\left|E_x\left(z,t\right)\right|^{2}\right)^{1/2}=\omega_0\sqrt{\frac{d}{\varepsilon_{0} w\ell}}q\left(t\right), \label{eq:v_res} \\
    i\left(t\right)=\left(\frac{\mu_0 w d}{L_{\text{eff}}}\int_{0}^{\ell}dz\left|H_y\left(z,t\right)\right|^{2}\right)^{1/2}=\pi \sqrt{\frac{w}{\mu_0 d\ell}}p\left(t\right). \label{eq:i_res}
\end{gather}
where we have used
\begin{equation}
    C_{\text{eff}}=\frac{\pi}{Z_0 \omega_0},\quad L_{\text{eff}}=\frac{Z_0}{\pi\omega_0}, \quad Z_0=\frac{d}{w}\sqrt{\frac{\mu_{0}}{\varepsilon_{0}}}.
\end{equation}  
in agreement with the definition in \cite{blais2021circuit}, where $Z_0$ is the characteristic impedance of the transmission line. Note that there is a degree of freedom in choosing $C_{\text{eff}}$ and $L_{\text{eff}}$. It determines the relation between the resonator impedance $Z = \sqrt{L_{\text{eff}}/C_{\text{eff}}}$ and $Z_0$, and is equivalent of changing our observation (coupling) point along the distributed resonator. 

After applying the second quantization on the fields, the annihilation operator (\ref{eq:annihilation_fields}) can be written as 
\begin{equation}
\hat{a}=\frac{1}{\sqrt{\hbar\omega_0}\sqrt{2Z\omega_0}}\left({\hat{v}\left(t\right)}+jZ\hat{i}\right).\label{eq:annihilation_z_independent}
\end{equation}
where $Z=Z_0/\pi$. Equation (\ref{eq:annihilation_z_independent}) is similar to its classical counterpart, (\ref{eq:12}), with the additional factor of $1/\sqrt{\hbar \omega_0}$ difference, as stated in \eqref{eq:classic_quantum}.

\section{Eigenmode analysis of coupled transmission lines} \label{sec:EMA_coupled_lines}

Consider a lossless reciprocal four-port network. Since the network is lossless,  
\begin{equation}
    [\mathbf{S}][\mathbf{S}^{*t}]=\mathbf{I}_{4 \times 4}
\end{equation}
where $\mathbf{S}^{*t}$ is the conjugate transpose of the $S$-matrix, and $\mathbf{I}$ is the identity matrix. Reciprocity also imposes \begin{equation}
    [\mathbf{S}]=[\mathbf{S}^{t}].
\end{equation} 
If zero reflection from all ports is also enforced (i.e. zero diagonal elements), the resulting $S$-matrix can always be reduced to either of the two forms (ports names may need adjustments) \cite{bahl2001rf} 
\begin{equation*}
    [\mathbf{S}]=\begin{bmatrix}
    0 & 0 & C_1 & C_2\\
    0 & 0 & C_2 & -C_1\\
    C_1 & C_2 & 0 & 0\\
    C_2 & -C_1 & 0 & 0
  \end{bmatrix} \quad \mathrm{or}
  \end{equation*}
  \begin{equation}
  [\mathbf{S}]=\begin{bmatrix}
    0 & C_1 & \pm j C_2 & 0 \\
    C_1 & 0 & 0 & \pm j C_2\\
    \pm j C_2 & 0 &0 & C_1\\
    0 & \pm j C_2 & C_1 & 0
  \end{bmatrix}. \label{eq:direction_coupler_S_matrix}
\end{equation}

Zero reflection from all four ports can be realized by either using the generalized S-matrix, or by impedance matching them to the common 50 $\Omega$ terminations. Here, we assume the latter. The resulting device, represented by (\ref{eq:direction_coupler_S_matrix}), is called a directional coupler since the input power to any port only exits from two ports. Note that the port numbers in (\ref{eq:direction_coupler_S_matrix}) are arbitrary, and the zero elements in each row are not necessarily next to each other.  
So far, we have only assumed zero loss and reciprocity for the four port network. Zero reflection from the ports in such networks automatically leads to a directional coupler device. 

Next, consider two identical parallel and uniformly coupled transmission lines represented by a reciprocal 4-port microwave network shown in Fig. \ref{fig:PTL}.

Because of the  symmetry, the S-matrix of the network can be written as 
\begin{equation}
  \mathbf{S =
    \begin{bmatrix}
    \mathbf{S}_{\mathrm{A}} & \mathbf{S}_{\mathrm{B}}\\
    \mathbf{S}_{\mathrm{B}} & \mathbf{S}_{\mathrm{A}}
  \end{bmatrix} \, ,
  \quad
  \mathbf{S}_{\mathrm{A}} =
  \begin{bmatrix}
    S_{11} & S_{12}\\
    S_{12} & S_{22}
  \end{bmatrix} \, ,
  \quad
  \mathbf{S}_{\mathrm{B}} =
  \begin{bmatrix}
    S_{31} & S_{41}\\
    S_{41} & S_{42}
  \end{bmatrix} \,} .
\end{equation}
Also, the symmetry requires the eigen modes of the coupled lines to be the even and odd modes. That is, the electric fields on the lines have equal intensity and zero or $\pi$ phase difference in even and odd modes, respectively. It can be shown that 
\begin{equation}
  \mathbf{S}_{\mathrm{A}}=\frac{\mathbf{S}_{\mathrm{+}}+\mathbf{S}_{\mathrm{-}}}{2}; \,\,\,\,\mathbf{S}_{\mathrm{B}}=\frac{\mathbf{S}_{\mathrm{+}}-\mathbf{S}_{\mathrm{-}}}{2}.  
\end{equation}
  
$\mathbf{S}_{\mathrm{+}}(\mathbf{S}_{\mathrm{-}})$ is the $S$-parameter of the two port network (ports 1,3 or 2,4) after placing a magnetic (electric) wall between the two transmission lines.  $\mathbf{S}_{\mathrm{+}}(\mathbf{S}_{\mathrm{-}})$ is also known as the even(odd) mode of the system. Reflections from the ports are 
\begin{equation}
    S_{11}=S_{22}=\frac{S_{\mathrm{11+}}+S_{11-}}{2};\,\,\,\,  S_{33}=S_{44}=\frac{S_{\mathrm{22+}}+S_{\mathrm{22-}}}{2}
\end{equation}

The forward-wave coupling (FC) and the reverse-wave coupling (RC) coefficients are defined as  
\begin{equation}
    \mathrm{FC}=\frac{\mathbf{S}_{\mathrm{21+}}+\mathbf{S}_{\mathrm{21-}}}{2},
\end{equation}
\begin{equation}
    \mathrm{RC}=\frac{S_{\mathrm{22+}}-S_{\mathrm{22-}}}{2}.
\end{equation}

In order to realize a directional coupler with zero reflections from the inputs, there are two convenient choices: 
\begin{enumerate}
    \item Forward-wave or co-directional coupling, which happens if $S_{\mathrm{11+}}=S_{11-}=S_{\mathrm{22+}}=S_{\mathrm{22-}}=0$. Equivalently,
\begin{equation}
    \beta_{\mathrm{+}}\neq\beta_{\mathrm{-}};\,\,\,\,Z_{\mathrm{+}}=Z_{\mathrm{-}}=Z_0
\end{equation} 
where $\beta_i$ are the propagation constants, $Z_i$ are the modes' characteristic impedances and $Z_0$ is the reference impedance for the $S$-parameters (i.e. terminations). The transferred power wave to the coupled line is 
\begin{equation}
    \left|S_{41}\right|=\sin\left(\frac{\left(\beta_{\mathrm{+}}-\beta_{\mathrm{-}}\right)l}{2}\right)
\end{equation}
This condition cannot be satisfied in transverse electromagnetic (TEM) transmission lines with homogeneous dielectrics because the phase velocities of the modes are equal. 
Note that a complete transfer of power to the coupled line is possible in forward-wave couplers. Also, there is always a 90 degrees phase difference between the coupled and direct line outputs. 
\begin{equation}
    \angle S_{41}-\angle S_{31}=90^o
\end{equation}
    \item  Backward-wave coupling, which happens if  $S_{\mathrm{11+}}=-S_{11-}$, $S_{\mathrm{22+}}=-S_{\mathrm{22-}}$, and $\mathbf{S}_{\mathrm{21+}}=S_{21_{\mathrm{-}}}$. 
Equivalently, 
\begin{equation}
    \beta_{\mathrm{+}}=\beta_{\mathrm{-}};\,\,\,\,Z_{\mathrm{+}}Z_{\mathrm{-}}=Z_0^2 \label{eq:backward_coupling_conditions}
\end{equation}
It can be shown that 
\begin{equation}
    S_{31}=\frac{\sqrt{1-\zeta^2}}{\sqrt{1-\zeta^2}\cos\theta+j\sin\theta},
\end{equation}
\begin{equation}
    S_{21}=\frac{j\zeta \sin\theta}{\sqrt{1-\zeta^2}\cos\theta+j\sin\theta},\label{eq:s31}
\end{equation}
where $\theta=\beta l$ is the electrical length, and $\zeta$ is the voltage coupling coefficient per $\theta$, when $\theta \rightarrow 0$, 
\begin{equation}
    \zeta=\frac{Z_{\mathrm{+}}-Z_{\mathrm{-}}}{Z_{\mathrm{+}}+Z_{\mathrm{-}}}.\label{eq:kappa_UL}
\end{equation}
Note that a complete transfer of power to the coupled line is impossible in this case. The phase difference between the outputs of the direct and coupled lines is still 90 degrees. 
Equations (\ref{eq:backward_coupling_conditions})-(\ref{eq:kappa_UL}) are very useful in multiplexing distributed resonators.

\end{enumerate}

Summarizing the useful relations,  
\begin{equation}
    Z_{\mathrm{+}}=\sqrt{\frac{L+L_m}{C-C_m}}; \quad Z_{\mathrm{-}}=\sqrt{\frac{L-L_m}{C+C_m}},
\end{equation}
	
\begin{equation}
    \omega L=\frac{\beta_{\mathrm{+}}Z_{\mathrm{+}}+\beta_{\mathrm{-}}Z_{\mathrm{-}}}{2};\quad  \omega L_m=\frac{\beta_{\mathrm{+}}Z_{\mathrm{+}}-\beta_{\mathrm{-}}Z_{\mathrm{-}}}{2},
\end{equation}		

\begin{equation}
    2\omega C=\frac{\beta_{\mathrm{-}}}{Z_{\mathrm{-}}}+\frac{\beta_{\mathrm{+}}}{Z_{\mathrm{+}}};\quad     2\omega C_m=\frac{\beta_{\mathrm{-}}}{Z_{\mathrm{-}}}-\frac{\beta_{\mathrm{+}}}{Z_{\mathrm{+}}}.
\end{equation}	

In a backward-wave directional coupler, 
\begin{equation}
    \beta_{\mathrm{+}}=\beta_{\mathrm{-}} \Rightarrow \,\,\frac{L_m}{L}=\frac{C_m}{C}. \label{eq:TL_balanced_condition}
\end{equation}

In  a forward-wave direction coupler,  
\begin{equation}
   Z_{\mathrm{+}}=Z_{\mathrm{-}} \Rightarrow \,\,\frac{L_m}{L}=-\frac{C_m}{C}.\label{eq:DC_impednace_condition_forwared}
\end{equation}

\section{Theory of weakly coupled transmission lines}\label{sec:Weakly_coupled_lines}

 This theory is limited to the forward-wave coupling between weakly coupled transmission lines. Its main application in superconducting devices is to calculate the cross-talk between TEM transmission lines. The theory assumes the following relations for the transmission lines voltages, 
 \begin{gather}
     \frac{dV_1}{dz}=-j\beta_1V_1-j\lambda V_2, \label{eq:dV1dz} \\
     \frac{dV_2}{dz}=-j\beta_2V_2-j\lambda V_1,\label{eq:dV2dz}
 \end{gather}
where the two transmission lines are along the z- axis with the coupling coefficient of $\lambda$, and the voltages and propagation constants of $V_{1,2}$ and $\beta_{1,2}$, respectively. By applying the initial condition $V_1=1$, $V_2=0$ at z=0,

\begin{equation*}
    V_1=\left[\frac{1}{2}+\frac{\beta_1-\beta_2}{2\sqrt{\left(\beta_1-\beta_2\right)^2+4\lambda^2}}\right]e^{-j\beta_s z}+\qquad \qquad
\end{equation*}
\begin{equation}
   \qquad \qquad \qquad \left[\frac{1}{2}-\frac{\beta_1-\beta_2}{2\sqrt{\left(\beta_1-\beta_2\right)^2+4\lambda^2}}\right]e^{-j\beta_f z},
\end{equation}

\begin{equation}
    V_2=\frac{\lambda}{2\sqrt{\left(\beta_1-\beta_2\right)^2+4\lambda^2}}\left(e^{-j\beta_s z}-e^{-j\beta_f z}\right),
\end{equation}
where $\beta_s=\frac{\beta_1+\beta_2}{2}+\frac{\sqrt{\left(\beta_1-\beta_2\right)^2+4\lambda^2}}{2}$ and $\beta_f=\frac{\beta_1+\beta_2}{2}-\frac{\sqrt{\left(\beta_1-\beta_2\right)^2+4\lambda^2}}{2}$ are usually called the slow and fast propagating coupled modes, respectively. In other words, in the presence of the coupling, slow and fast waves are excited and their interference determines the power distribution on the two lines along the propagation direction. If the lines are symmetrical, $\beta_1=\beta_2=\beta_0$,
\begin{gather}
    V_1=\cos\left(\lambda z\right)e^{-j\beta_0z},\label{eq:weakly_coupled_lines_V1} \\
    V_2=-j\sin\left(\lambda z\right)e^{-j\beta_0z}, \\
    \lambda=\frac{\beta_s-\beta_f}{2}. \label{eq:weakly_coupled_lines_coupling}
\end{gather}

Equations (\ref{eq:weakly_coupled_lines_V1})-(\ref{eq:weakly_coupled_lines_coupling})  are consistent with the forward-wave directional coupler relations, extracted in the previous section. They can be used to extract the coupling between transmission lines from the propagating eigen modes.  

A more physical description of this theory can also be reviewed by considering the fields instead of voltages \cite{okamoto2021fundamentals,lecture_notes}. Consider two parallel transmission lines along the z- axis. The transmission lines support the bare modes of $\Vec{E_1}\left(x,y\right)e^{-j\beta_1z}$ and $\Vec{E_2}\left(x,y\right)e^{-j\beta_2z}$ in isolation. Let us define a ``super-mode'' as the sum of the bare modes with z- dependent coefficients (assuming the weak coupling does not change the bare modes dramatically) as  
\begin{equation}
    \Vec{E}\left(x,y,z\right)=A\left(z\right)\Vec{E}_1\left(x,y\right)e^{-j\beta_1z}+B\left(z\right)\Vec{E}_2\left(x,y\right)e^{-j\beta_2z}
\end{equation}
The same coefficients apply to the magnetic field of the super-mode. It can be shown that $A(z)$ and $B(z)$ must satisfy the following conditions (known as generalized coupled mode equations): 
\begin{equation}
\frac{dA}{dz}+c_{12}\frac{dB}{dz}e^{-j\left(\beta_{2}-\beta_{1}\right)z}+j\beta_{1}A+j\lambda_{12}Be^{-j\left(\beta_{2}-\beta_{1}\right)z}=0
\end{equation}
\begin{equation}
\frac{dB}{dz}+c_{21}\frac{dA}{dz}e^{-j\left(\beta_{2}-\beta_{1}\right)z}+j\beta_{2}A+j\lambda_{21}Ae^{-j\left(\beta_{2}-\beta_{1}\right)z}=0
\end{equation}
in which,
\begin{equation}
    \lambda_{12}=\frac{\omega\varepsilon_{0}\iint_{\infty}^{\infty}ds\left(\varepsilon_{r}-\varepsilon_{r,2}\right){\Vec{E}_{1}^{*}\cdot \Vec{E}_{2}}}{\iint_{\infty}^{\infty}ds\,\hat{z}\cdot\left({\Vec{E}_{1}^{*}\times \Vec{H}_{1}+\Vec{E}_{1}\times \Vec{H}_{1}^{*}}\right)}
\end{equation}
is the coupling coefficient and measure of power leakage from one transmission line to the other one, and $\varepsilon_{r,2}$ is the dielectric function with only transmission line 1.  The term $\left(\varepsilon_r-\varepsilon_{r,2}\right)$ means that we only consider transmission line 1 for the dielectric function. The integration is over the cross section of the transmission lines. Also,
\begin{equation}
    c_{12}=\frac{\iint_{\infty}^{\infty}ds\,\hat{z}\cdot\left({\Vec{E}_{1}^{*}\times \Vec{H}_{2}+\Vec{E}_{2}\times \Vec{H}_{1}^{*}}\right)}{\iint_{\infty}^{\infty}ds\,\hat{z}\cdot\left({{\Vec{E}_{1}^{*}\times \Vec{H}_{1}+\Vec{E}_{1}\times \Vec{H}_{1}^{*}}}\right)}
\end{equation}
is the excitation efficiency. It quantifies the power fed to the unexcited transmission line by the excited transmission line, at the input. The change in the propagation constant of the transmission line 1, due to the presence of line 2, is 
\begin{equation}
    \beta_{1}=\frac{\omega\varepsilon_{0}\iint_{\infty}^{\infty}ds\,\left(\varepsilon_{r}-\varepsilon_{r.2}\right){\Vec{E}_{1}^{*}\cdot \Vec{E}_{1}}}{\iint_{\infty}^{\infty}ds\,\hat{z}\cdot\left({{\Vec{E}_{1}^{*}\times \Vec{H}_{1}+\Vec{E}_{1}\times \Vec{H}_{1}^{*}}}\right)}.
\end{equation}

Ignoring the excitation coupling, and assuming $\beta_1\simeq\beta_2$ and reciprocity,   

\begin{equation}
\frac{dA}{dz}=-j\beta B-j\lambda A, \label{eq:dAdz}
\end{equation}

\begin{equation}
\frac{dB}{dz}=-j\beta A-j\lambda B,\label{eq:dBdz}
\end{equation}
which are similar to (\ref{eq:dV1dz}) and (\ref{eq:dV2dz}). The super-mode propagates as 
\begin{equation}
E=\left[E_{1}\left(x,y\right)\cos\left(\left|\kappa_{12}\right|z\right)+E_{2}\left(x,y\right)\sin\left(\left|\kappa_{12}\right|z\right)\right]e^{-j\beta z}.
\end{equation}

In other words, the coupling between transmission lines grows with length, and there is a complete transfer of power from one transmission line to the other at $z={\pi}/\left({2\left|\lambda\right|}\right)$. For lengths much smaller than  $z={\pi}/\left({2\left|\lambda\right|}\right)$,
\begin{equation}
\frac{P_{2}\left(x\right)}{P_{1}\left(x\right)}=\sin^{2}\left(\left|\lambda \right|z\right)\simeq\left|\lambda\right|^{2}z^{2}.
\end{equation}

Based on the eigen mode analysis results, $\beta_1=\beta_2$ along with $Z_1Z_2=Z_0^2$ prevent forward-wave coupling in the geometry. This means $Z_1Z_2=Z_0^2$ must lead to $\lambda=0$. In obtaining (\ref{eq:dAdz}) and (\ref{eq:dBdz}), we assumed $\beta_1\simeq\beta_2$, but they cannot be exactly equal (i.e. $\beta_1\neq\beta_2$.) For additional references, see \cite{haus1991coupled,marcatili1986improved,pierce1954coupling,schelkunoff1955conversion,yariv1973coupled}.

\section{Eigenmode analysis of backward coupler}

For sake of completion, let us verify the loss rate and scattering of a quarter-wave resonator coupled to a transmission line. The eigenmode of the circuit can be found by computing the solutions to $\text{det}(Y) = 0$ where $Y$ is the admittance. The admittance of the coupler is given by
\begin{equation}
    Y_{BC} = \frac{j Y_0}{\sqrt{1-\zeta^2}} \begin{pmatrix}
    -\cot\theta & \csc\theta \\ \csc\theta & -\cot\theta
    \end{pmatrix} \otimes 
    \begin{pmatrix} 1 & -\zeta \\ -\zeta & 1 \end{pmatrix},
\end{equation}
where $Y_0 = 1/Z_0$. To the coupler we add the following admittance matrix to replicate the scenario in Fig. \ref{fig:coupler_phase_matched}:
\begin{equation}
    Y_\Gamma = Y_0 \begin{pmatrix} 1 & 0 & 0 & 0 \\ 0 & -j \cot\beta \ell_1 & 0 & 0 \\ 0 & 0  & j \tan\beta\ell_f & 0 \\ 0 & 0 & 0 & j \tan\beta\ell_2 \end{pmatrix},
\end{equation}
which uses a matched port for port 1. The parameter $\ell_f$ is the length of the open termination on the feedline while $\ell_1$ and $\ell_2$ comprise a $\lambda/4$ resonator as shorted and open terminations, respectively. In the weak coupling limit $\zeta^2 \ll 1$, we find a root corresponding to the resonator mode with
\begin{gather}
    \omega_r \approx \frac{\pi v_{ph}}{2 \ell_r}\left[1 + \zeta^2 \left(\frac{\ell_c}{2\ell_r} - \frac{\sin(\frac{\pi \ell_c}{\ell_r})}{2\pi}\right)\right], \\
    \kappa_r \approx \frac{2\zeta^2 v_{ph}}{\ell_r} \sin^2\left(\frac{\pi \ell_c}{2\ell_r}\right),
\end{gather}
for frequency and decay rate, respectively, where $\ell_r = \ell_1 + \ell_2 + \ell_c$ is the resonator length, $\ell_c$ is the coupler length and we have taken $\ell_f = \ell_1 + \ell_r$ to maximize $\kappa_r$. The reflection coefficient can be found by contracting the scattering matrix on ports 2-4. In the weak coupling limit and near resonance, it can be shown that
\begin{align}
    S_{11} \approx e^{-j 2 \beta (\ell_c + \ell_f)} \frac{\kappa_r/2 - j(\omega - \omega_r)}{\kappa_r/2 + j(\omega - \omega_r)},
\end{align}
which takes the standard expected form for a resonant object read out in reflection.

\section{Coupled resonators: alternative formulation}\label{sec:coupled_res_symmetric}

Let us re-define resonance mode amplitudes as 

\begin{align}
    a_{\pm} &= \frac{1}{\sqrt{2  Z }}\left(v\pm j Z i\right),  \label{eq:a_no_omega}
\end{align}
such that the energy in a corresponding uncoupled resonator becomes 
\begin{equation}
    W=\frac{a_+a_-}{\omega_0}.\label{eq:W_no_omega}
\end{equation}
Then using \eqref{eq:a_no_omega}, \eqref{eq:9}, and \eqref{eq:10}, 
\begin{gather}
    \dot{a}_{k\pm}=\pm j\omega_{0k}a_{k\pm}+\sqrt{\frac{\omega_{0k}}{\omega_{0l}}}\frac{\left(\zeta_{C}\mp\zeta_{L}\right)}{2}\dot{a}_{l+}+\sqrt{\frac{\omega_{0k}}{\omega_{0l}}}\frac{\left(\zeta_{C}\pm\zeta_{L}\right)}{2}\dot{a}_{l-}, \label{eq:coupled_resonators_no_omega}
\end{gather}
Recasting (\ref{eq:coupled_resonators_no_omega}) to isolate derivatives,

\begin{equation}
    \begin{pmatrix}
    \dot{\mathbf{a}}_{1}/\sqrt{\omega_{01}}\\
    \dot{\mathbf{a}}_{2}/\sqrt{\omega_{02}}
    \end{pmatrix} = \begin{pmatrix}
    \mathbf{U}_0 & \mathbf{U}_g
    \\
    \mathbf{U}_g & \mathbf{U}_0 \end{pmatrix} \begin{pmatrix}
    j\sqrt{\omega_{01}}\mathbf{a}_{1}\\
    j\sqrt{\omega_{02}}\mathbf{a}_{2}
    \end{pmatrix} \label{eq:matrix_coupled_res_no_omega}
\end{equation}
maintaining definitions \eqref{eq:matrix_coupled_res}, \eqref{eq:coupled_res_matrix_defs}, and \eqref{eq:coupled_res_constants}. 
Following the small coupling limit and RWA, we get
\begin{equation}
    \dot{a}_{k\pm} \approx \pm j \omega_{0k} a_{k\pm} \pm j \sqrt{\omega_{0l}\omega_{0k}} \frac{\zeta_{C}-\zeta_{L}}{2} a_{l\pm}, \label{eq:coupled_res_approximate2_no_omega}
\end{equation}
Then, the total energy in the system is 
\begin{align}
    W_{\text{tot}} = & \frac{a_{1+}a_{1-}}{\omega_{01}} + \frac{a_{2+}a_{2-}}{\omega_{02}} -\frac{\zeta_C}{2\sqrt{\omega_{01}\omega_{02}}}\left(a_{1+} + a_{1-}\right)\left(a_{2+} + a_{2-}\right) \nonumber \\
    & -\frac{\zeta_L}{2\sqrt{\omega_{01}\omega_{02}}}\left(a_{1+} - a_{1-}\right)\left(a_{2+} - a_{2-}\right). \label{eq:W_tot_no_omega}
\end{align}
Equation \eqref{eq:coupled_resonators_no_omega} naturally satisfies energy conservation $dW_{\text{tot}}/dt = 0$.
Note that the coupling term in the right hand side of \eqref{eq:coupled_res_approximate2_no_omega} is the same for both equations. This form is only obtained by choosing the definitions in \eqref{eq:a_no_omega} and \eqref{eq:W_no_omega}.

\bibliographystyle{IEEEtran}
\bibliography{Couplings}


\end{document}